\documentstyle[epsfig,aps]{revtex}
\newcommand{\nc}{\newcommand}
\nc{\rnc}{\renewcommand}
\nc{\acs}{\arraycolsep}
\nc{\mc}{\multicolumn}
\nc{\bsk}{\baselineskip}
\nc{\vsp}{\vspace}
\nc{\hsp}{\hspace}
\nc{\stl}{\setlength}
\nc{\stc}{\setcounter}
\nc{\addl}{\addtolength}
\nc{\beq}{\begin{equation}}
\nc{\eeq}{\end{equation}}
\nc{\beqa}{\begin{eqnarray}}
\nc{\eeqa}{\end{eqnarray}}
\nc{\tfrac}[2]{\raisebox{.4ex}{\tiny $\frac{#1}{#2}$}} 
\nc{\romlist}{ \setcounter{num1}{0}%
  \begin{list}{(\roman{num1})}{\usecounter{num1}} }
\nc{\arblist}{ \setcounter{num1}{0}%
  \begin{list}{(\arabic{num1})}{\usecounter{num1}} }
\nc{\alphlist}{ \setcounter{num2}{0}%
  \begin{list}{(\alph{num2})}{\usecounter{num2}} }
\nc{\bullist}{\begin{list}{$\bullet$}{ }}
\nc{\nr}{\\ \hline}
\nc{\hrl}{{\center \stl{\unitlength}{\textwidth} 
 \begin{picture}(1,0)  \put(0,0){\line(1,0){1}}
 \end{picture} \vsp{.001\bsk} }}
\nc{\cents}{{\scriptsize$\mbox{\rm C}\!\!\!\mbox{\raisebox{.2ex}%
{$|$}}\,\,\,\,$}}
\nc{\figsp}[5]{\begin{figure}[#1] \vsp{#2} \caption[#4]{#3} 
\label{#5} \vsp{2\bsk} \end{figure}}
\nc{\fig}{\figsp{tbp}}
\nc{\figb}{\figsp{b}}
\nc{\figh}{\figsp{h}}
\nc{\llist}{\begin{list}{}{} \stl{\labelsep}{.4in}}
\nc{\lit}[2]{
 \item[\raggedright #1]{#2}}
\nc{\lbit}[2]{ 
 \item[\raggedright\bf #1]{#2}}
\nc{\lemit}[2]{ 
 \item[\raggedright\em #1]{#2}}
\nc{\lbemit}[2]{ 
 \item[\raggedright\bf\em #1]{#2}}
\nc{\clst}[1]{\stl{\coltwo}{\textwidth}
\addl{\coltwo}{-#1} \addl{\coltwo}{-5.56ex} \newline
\begin{tabular}{p{#1}p{\coltwo}} \citem{}{}}
 
\nc{\citem}[2]{{\raggedright \bf #1} & #2 \\ }
\nc{\cemitem}[2]{{\raggedright \em #1} & #2 \\ }
\nc{\cbemitem}[2]{{\raggedright \bf \em #1} & #2 \\ }
\nc{\cend}{\citem{}{} \end{tabular} 
\mbox{}
} 
\nc{\SSP}{{\rm \hsp{.4in}}}
\nc{\SSPP}{{\rm \hsp{.2in}}}
\nc{\ds}{\displaystyle}
\nc{\tx}{\textstyle}
\nc{\scst}{\scriptstyle}
\nc{\sscst}{\scriptscriptstyle}
\nc{\prt}{\partial}
\nc{\fr}{\frac}
\nc{\lf}{\left}
\nc{\rt}{\right}
\nc{\la}{\langle}
\nc{\ra}{\rangle}
\nc{\V}{\vec}
\nc{\str}{\stackrel}
\nc{\ovl}{\overline}
\nc{\ul}{\underline}
\nc{\ovb}{\overbrace}
\nc{\ub}{\underbrace}
\nc{\wh}{\widehat}
\nc{\B}{\bar}
\nc{\D}{\dot}
\nc{\C}{\cdot}
\nc{\dd}{\ddot}
\nc{\tl}{\tilde}
\nc{\ha}{\hat}
\nc{\nn}{\nonumber}
\nc{\app}{\approx}
\nc{\al}{\alpha}
\nc{\RA}{\rightarrow}
\nc{\LRA}{\leftrightarrow}
\nc{\SRA}{\SSP\rightarrow\SSP}
\nc{\SSRA}{\SSPP\rightarrow\SSPP}
\nc{\dg}{\dagger}
\nc{\vp}{\varphi}
\nc{\ve}{\varepsilon}
\nc{\Dl}{\Delta}
\nc{\dl}{\delta}
\nc{\gm}{\gamma}
\nc{\Gm}{\Gamma}
\nc{\ep}{\epsilon}
\nc{\sg}{\sigma}
\nc{\Sg}{\Sigma}
\nc{\ua}{\uparrow}
\nc{\da}{\downarrow}
\nc{\lam}{\lambda}
\nc{\eql}[1]{\parbox{#1\textwidth}}
\nc{\eqm}[1]{\makebox[#1\textwidth][l]}
\nc{\enu}[1]{\mbox{\hspace{.4in}(\theequation.#1)}}
\nc{\son}{\\ \\ \ds}
\nc{\stw}{\\ & \\ \ds}		%   Equation formatting   %
\nc{\sth}{\\ & & \\ \ds}
\nc{\sfo}{\\ & & & \\ \ds}
\nc{\sfi}{\\ & & & & \\ \ds}
\nc{\A}{& \ds}
\nc{\bbr}{\lf\{\rule[-1.5ex]{0in}{0.01in}\rt.}
\nc{\hf}{\fr{1}{2}}
\nc{\mhf}{\mbox{\footnotesize$\hf$}}
\nc{\dv}{\/!}
\nc{\dint}{\int\!\!\int}
\nc{\tint}{\int\!\!\dint}               % integrals %
\nc{\qint}{\int\!\!\tint}
\nc{\Pd}[2]{\fr{\prt #1}{\prt #2}}
\nc{\Pdt}[1]{\Pd{#1}{t}}
\nc{\Pdx}[1]{\Pd{#1}{x}}
\nc{\Pdy}[1]{\Pd{#1}{y}}
\nc{\Pdz}[1]{\Pd{#1}{z}}           	%  Derivatives  %           
\nc{\Pdr}[1]{\Pd{#1}{r}}
\nc{\Pds}[1]{\Pd{#1}{s}}
\nc{\Dv}[2]{\fr{d#1}{d#2}}
\nc{\Dvt}[1]{\Dv{#1}{t}}
\nc{\Dvx}[1]{\Dv{#1}{x}}
\nc{\Dvy}[1]{\Dv{#1}{y}}
\nc{\Dvz}[1]{\Dv{#1}{z}}
\nc{\Dvr}[1]{\Dv{#1}{r}}
\nc{\Drs}[1]{\Dv{#1}{s}}
\nc{\inpp}[3]{\la #1| #2| #3\ra}
\nc{\inp}[2]{\inpp{#1}{#2}{#1}}
\nc{\rb}[1]{| #1\ra}
\nc{\lb}[1]{\la#1|}
\nc{\dtpp}[2]{\lb{#1}\rb{#2}}		%  Inner Products  %
\nc{\dtp}[1]{\dtpp{#1}{#1}}
\nc{\otpp}[2]{\rb{#1}\lb{#2}}
\nc{\otp}[1]{\otpp{#1}{#1}}
\newcounter{num1} \newcounter{num2}  %for \romlist and \alphlist
\newlength{\coltwo}
\rnc{\L}{{\cal L}}                      %  Miscellaneous  %
\nc{\lapp}{\mbox{\raisebox{-.6ex}{$\,\stackrel{\textstyle <}{\sim}\,$}}}
\nc{\gapp}{\mbox{\raisebox{-.6ex}{$\,\stackrel{\textstyle >}{\sim}\,$}}}
\nc{\als}{\fr{\al_s(Q^2)}{2\pi}}
\nc{\gpx}{g_1^p(x,Q^2)}
\nc{\gpz}{g_1^p(z,Q^2)}
\nc{\muq}{\lf(\fr{\mu^2}{Q^2}\rt)}
\nc{\xy}{(\fr{x}{y})}
\nc{\ASQ}{\al_s(Q^2)}
\nc{\Li}{{\rm Li}_2}
\nc{\dqx}{\Dl q_i(x,Q^2)} \nc{\dqy}{\Dl q_i(y,Q^2)} 
\nc{\dQ}{\Dl q_i(Q^2)}
\nc{\dgx}{\Dl g(x,Q^2)} \nc{\dgy}{\Dl g(y,Q^2)} 
\nc{\dG}{\Dl g(Q^2)}
\nc{\xq}{(x,Q^2)} \nc{\yq}{(y,Q^2)}
\nc{\Tt}{\tl{t}} \nc{\Ts}{\tl{s}} \nc{\Tu}{\tl{u}}
\nc{\Hs}{\ha{s}} \nc{\Ht}{\ha{t}} \nc{\Hu}{\ha{u}}
\nc{\Hsg}{\hat{\sg}}
\nc{\GeV}{\mbox{\rm GeV}}
\nc{\sS}{\!\not{\!s}}  
\nc{\pS}{\!\not{\!p}}  \nc{\kS}{\!\not{\!k}}
\nc{\poS}{\!\not{\!p}_1}  \nc{\pwS}{\!\not{\!p}_2}
\nc{\ptS}{\!\not{\!p}_3}  \nc{\pfS}{\!\not{\!p}_4}
\nc{\AS}{\!\not{\!\!A}}  \nc{\ASS}{\!\not{\!\!A}^*}
\nc{\BS}{\!\not{\!\!B}}  \nc{\BSS}{\!\not{\!\!B}^*}
\nc{\Tr}{\mbox{\rm Tr}}
\nc{\pT}{p_T}
\nc{\xT}{x_T}
\nc{\AoS}{\!\not{\!\!A}_1}  \nc{\AoSS}{\!\not{\!\!A}_1^*}
\nc{\AwS}{\!\not{\!\!A}_2}  \nc{\AwSS}{\!\not{\!\!A}_2^*}
\nc{\BoS}{\!\not{\!\!B}_1}  \nc{\BoSS}{\!\not{\!\!B}_1^*}
\nc{\BwS}{\!\not{\!\!B}_2}  \nc{\BwSS}{\!\not{\!\!B}_2^*}
\nc{\aS}{\!\not{\!a}}  \nc{\bS}{\!\not{\!b}}
\nc{\aoS}{\!\not{\!a}_1}  \nc{\boS}{\!\not{\!b}_1}
\nc{\awS}{\!\not{\!a}_2}  \nc{\bwS}{\!\not{\!b}_2}
\nc{\anS}{\!\not{\!a}_n}  \nc{\bnS}{\!\not{\!b}_n}
\nc{\anpoS}{\!\not{\!a}_{n+1}}  \nc{\bnpoS}{\!\not{\!b}_{n+1}}
\nc{\anmoS}{\!\not{\!a}_{n-1}}  \nc{\bnmoS}{\!\not{\!b}_{n-1}}
\nc{\refi}[1]{$^{\,\mbox{\scriptsize \ref{#1}}}$}
\nc{\refii}[2]{$^{\,\mbox{\scriptsize \ref{#1},\ref{#2}}}$}
\nc{\refiii}[3]{$^{\,\mbox{\scriptsize \ref{#1},\ref{#2},\ref{#3}}}$}
\nc{\refr}[2]{$^{\,\mbox{\scriptsize \ref{#1}--\ref{#2}}}$}
\nc{\sint}{\int \!\!}
\nc{\qB}{\stackrel{(-)}{q}}
\nc{\Lam}{\Lambda}

\begin{document}

\draft

%\stl{\bsk}{1.5\bsk}
%\pagestyle{empty}

\title{Analytically expanded and integrated results for massive fermion
production in two-photon collisions and a high precision
$\al_s$ determination}

\author{B.\ Kamal}

\address{Physics Department, Brookhaven National Laboratory, Upton,
 New York 11973, U.S.A.}

\author{Z.\ Merebashvili$^*$}
\address{Physics Department, McGill University, Montreal,
 Quebec H3A 2T8, Canada}

\date{March 1998}

\maketitle

\widetext

\begin{abstract}
The  cross section for massive fermion production in
two-photon collisions was examined at next-to-leading order in
QCD/QED for general photon helicity.
The delta function (virtual+soft) part of the differential
cross section was analytically integrated over the final state phase
space. Series expansions for the complete differential and total cross
sections were given up to tenth order in the parameter $\beta$. These
were shown to be of practical use and revealed much structure.
Accurate parametrizations of the total cross sections were given, valid
up to higher energies. The above results were applied to top quark
production in the region not too far above threshold. The cross section
was shown to be quite sensitive to $\alpha_s$ in the appropriate
energy region.
\end{abstract}

%\vsp{.15in}
%\noindent

%\narrowtext

\renewcommand{\thefootnote}{\fnsymbol{footnote}}
\addtocounter{footnote}{1}
\footnotetext{Present address: High Energy Physics Institute, Tbilisi
State University, University st.\ 9, 380086
Tbilisi, Republic of Georgia.}

\nc{\tlo}[1]{\stackrel{(\sim)}{#1}}
\nc{\omg}{\omega}
\nc{\Lit}{{\rm Li}_2}
\nc{\lih}{{\rm Li}_3}

%\vglue .3cm
%\begin{center}\begin{large}\begin{bf}
\section{INTRODUCTION}
%\end{bf}\end{large}\end{center}
%\vglue .3cm

High energy photons may be produced by backscattering laser light off
high energy $e^-$ or $e^+$ beams. In addition, high degrees of polarization
are possible and the photons may carry a large fraction of the
electron energy. Photon-photon collisions also arise naturally as a
background in $e^+e^-$ collisions. One major motivation for constructing
a $\gm\gm$ interaction region at a high energy next linear collider (NLC)
is to produce Higgs bosons on
resonance via $\gm\gm$ fusion, which also allows direct determination of
the $H\gm\gm$ coupling, which is sensitive to possible non Standard 
Model charged particles
of large mass that may enter in the triangle loop.
Using polarized photons allows one to control the backgrounds arising
from $\gm\gm\RA b\B{b}$, for an intermediate mass Higgs \cite{Gun}. 
This background has
now been studied including QCD \cite{heavy,Jik,Bord,Fadin} and 
electroweak \cite{Denn} corrections.

In this paper we will consider in some detail the process 
$\gm\gm \RA f\B{f} + X$ in the region not too far above threshold, 
making use of the complete analytical results presented in \cite{heavy},
which include photon polarization. We will demonstrate the usefulness
of $\gm\gm \RA t\B{t} + X$ in determining $\alpha_s$ precisely.
We extend the analytical results presented in \cite{heavy} by 
integrating and obtaining analytical results for
the single integral (virtual+soft) part and by series
expanding the entire differential and integrated cross section
to order $\beta^{10}$, where $\beta$ is the massive fermion velocity
in the soft radiation limit. Such an expansion is shown to be of 
practical use, not too far above threshold, and it demonstrates many
interesting features of the corrected cross sections. We have also 
provided parametrizations of the total integrated cross sections valid
up to higher energies.

As the diagrams involve only QED-like vertices, the process under 
consideration is quite fundamental in nature. The fact that complete
analytical results have been absent, until recently, reflects the 
lack of experimental feasibility of directly colliding photons of 
high energy, although, as mentioned earlier, such collisions naturally
arise as a background 
in $e^+e^-$ collisions. It also reflects the
difficulty in obtaining and presenting in a compact form complete 
analytical results, including bremsstrahlung, when massive fermions
are present. The task would be even more formidable for reactions
such as $gg\RA Q\ovl{Q}+X$ \cite{Nas,Been}.
Our hope is that the approach of using high order series
expansions to simplify, and clarify, such results will become more
widespread. As well, our analytical integration of the single integral
part is an essential part of a complete analytical integration, which
is likely to be performed sometime in the near future. In the 
meantime, our parametrizations provide sufficient accuracy to be
useful, as do our series expansions closer to threshold.

The possibility of directly producing fermion pairs in $\gm\gm$ collisions
has recently been realized at SLAC \cite{SLAC}, where a high energy 
($\lesssim$ 29.2 GeV) photon beam was collided on a low energy (2.35 eV) 
laser beam.
Since the center of mass energy was insufficient to produce a pair
($e^+e^-$, in this case), multiple photon fusion was required;
a different mech--
\twocolumn \noindent 
anism than that being considered here. The high energy
beam was produced via backscattering (of the same 2.35 eV beam)
off a 46.6 GeV electron beam and represents a positive first
step towards the construction of a higher energy $\gm\gm$ collider
with both beams produced via backscattering. Of course, many technical
difficulties arise in such a machine and these have been investigated
(see \cite{Brink}).

%\vglue 1cm
%\begin{center}\begin{large}\begin{bf}
\section{GENERAL FORM AND DECOMPOSITION OF THE DIFFERENTIAL CROSS SECTION}
%\end{bf}\end{large}\end{center}
%\vglue .3cm

The process under consideration is

\beq
\gm(p_1,\lam_1) + \gm(p_2,\lam_2) \RA f(p_3) + \B{f}(p_4) + [V(k)],
\eeq
where $\lam_1$, $\lam_2$ denote helicities and the $p_i$, $k$ denote momenta.
$f$ ($=q,l$) represents a fermion with mass $m$ and $V=g,\gm$. The
square brackets represent the fact that there may or may not be a
gluon/photon in the final state. We have the following invariants,
\beqa
\nn
s &\equiv& (p_1+p_2)^2, \SSPP t \equiv T-m^2 \equiv 
(p_1-p_3)^2 - m^2, \\
 u &\equiv& U-m^2 \equiv (p_2-p_3)^2 -m^2
\eeqa
and
\beq
s_2 \equiv S_2 - m^2 \equiv (p_1+p_2-p_3)^2 - m^2 = s+t+u.
\eeq
Defining
\beqa
\nn
v &\equiv& 1+\frac{t}{s}, \SSP w \equiv \frac{-u}{s+t}, \\
 \beta &\equiv& 
\sqrt{1-4m^2/s}, \SSP x\equiv \fr{1-\beta}{1+\beta},
\eeqa
we may express
\beqa
\nn
t &=& -s(1-v), \SSP u = -svw, \\
  s_2 &=& sv(1-w), \SSP  m^2=\fr{s}{4}(1-\beta^2).
\eeqa
Now introduce
\beq \label{eg6}
\kappa(s) \equiv 2\pi\fr{\al^2e_f^4[N_c]}{s}, \SSPP C_1\equiv 
[C_F] \fr{\al_V}{2\pi},
\eeq
where the $N_c$, $C_F$ factors are present only for $f$ = quark and
$V$ = gluon, respectively and $e_f$ is the fermion's fractional charge.
Here
\beq
\alpha_V = \lf\{ 
\begin{array}{ll}
\alpha_s,  & V=g \\
\alpha,  & V=\gamma 
\end{array} 
\rt.\,\,.
\eeq
Then
\beqa
\fr{d\sg}{dvdw} &=& \fr{d\sg^{(0)}}{dvdw} + \fr{d\sg^{(1)}}{dvdw} \\
& \equiv& \kappa(s)\lf[ \fr{1}{2\pi}\fr{df^{(0)}}{dvdw} 
+ \fr{C_1}{\pi}\fr{df^{(1)}}{dvdw} \rt].
\eeqa
The $f$ functions are dimensionless functions of $v$ and $w$, which
allow us to parametrize our cross sections in an exact fashion, without
dependence on $\alpha_V$. 
We use the normalization convention of \cite{Kuhn}. 
Since, in that normalization, the $f^{(i)}$ contain an overall factor
of $\pi$, we consistently present analytical results for $f^{(i)}/\pi$
in order to cancel it.
The unpolarized and polarized $f^{(i)}$ are given by
\beqa
\nn
f^{(i)}_{\rm unp} &=& \fr{1}{2} [f^{(i)}(+,+) + f^{(i)}(+,-)],
\\
f^{(i)}_{\rm pol} &=& \fr{1}{2} [f^{(i)}(+,+) - f^{(i)}(+,-)],
\eeqa
in the notation $f^{(i)}(\lam_1,\lam_2)$. Define
\beq
j \equiv 1 - <\lam_1\lam_2>,
\eeq
where $<\lam_1\lam_2>$ is the average value of $\lam_1\lam_2$. Then
\beqa
\nn
f^{(i)}(j) &=& \fr{1+<\lam_1\lam_2>}{2} f^{(i)}(+,+) \\
& & + \fr{1-<\lam_1\lam_2>}{2} f^{(i)}(+,-) \\
\nn
&=& f^{(i)}(+,+) + \fr{j}{2}[f^{(i)}(+,-)-f^{(i)}(+,+)] \\
&=& f^{(i)}(+,+) - j f^{(i)}_{\rm pol},
\eeqa
so that
\beq
j = \lf\{ 
\begin{array}{ccl}
0 &\RA& f^{(i)}(j) = f^{(i)}(+,+) = f^{(i)}(-,-) \\
2 &\RA& f^{(i)}(j) = f^{(i)}(+,-) = f^{(i)}(-,+) \\
1 &\RA& f^{(i)}(j) = f^{(i)}_{\rm unp}
\end{array} 
\rt. \,\, .
\eeq
The LO term is given by
\beqa
\fr{1}{2\pi}\fr{df^{(0)}(j)}{dvdw} &=&\dl(1-w) \Biggl\{
\fr{2m^2/s}{v^2(1-v)^2} (1-2m^2/s) \\ \nn
&&  + j\lf(\fr{1}{v(1-v)}-2\rt)\lf[
1 - \fr{2m^2}{sv(1-v)}\rt]
\Biggr\} \\
\label{eb16}
&=&\dl(1-w) \Biggl\{ \fr{1-\beta^4}{4v^2(1-v)^2} \\ \nn
&&  + j\lf(\fr{1}{v(1-v)}-2\rt)\lf[ 
1 - \fr{1-\beta^2}{2v(1-v)}\rt]
\Biggr\}.
\eeqa
The last form shows explicitly the polynomial structure of the leading
order differential cross section in terms of $\beta$.
This is somewhat misleading, however, as we shall see in the next section,
since the phase space in $v$ itself depends on $\beta$.

From \cite{heavy} we see that $df^{(1)}/dvdw$ has the form
\beqa
\fr{1}{\pi}\fr{df^{(1)}}{dvdw} &=& F_h(v,w) + \fr{F_s(v,w)}{(1-w)_+}
+ F_\dl(v) \dl(1-w) \\
\nn
&=& F_h(v,w) + \fr{F_s(v,w)-F_s(v,1)}{1-w} \\
\nn & & + F_s(v,1) \ln(1-w_1)\dl(1-w) \\
\nn
& & + F_\dl(v) \dl(1-w)
+ \fr{1}{\pi}\fr{d\ha{f}^{(1)}}{dvdw} \\
&\equiv &  \fr{1}{\pi}\fr{d\tl{f}^{(1)}}{dvdw}
+ \fr{1}{\pi}\fr{d\ha{f}^{(1)}}{dvdw}  ,
\eeqa
where
\beq
\label{e19}
w_1 = \fr{1-\beta^2}{4 v (1-v)}
\eeq
and
\beqa
\fr{1}{\pi}\fr{d\ha{f}^{(1)}}{dvdw} = \fr{F_s(v,w)}{(1-w)_+}
- \biggl[ \fr{F_s(v,w)-F_s(v,1)}{1-w} \\
\nn 
+ F_s(v,1) \ln(1-w_1)\dl(1-w) \biggr].
\eeqa
When integrating from $w_1$ to 1, $d\ha{f}^{(1)}/dvdw$ makes no 
contribution, otherwise it contributes.
The function $F_\dl(v)\dl(1-w)$ has the form
\beqa
F_\dl(v)\dl(1-w) &=& F_s(v,1)\ln\lf(\fr{sv}{m^2}\rt)\dl(1-w) \\
\nn && + \fr{1}{2\pi}\fr{df^{(0)}}{dvdw} F_\dl^B(\beta)
 + F_\dl^{NB}(v)\dl(1-w)
\eeqa
and $F_s(v,1)$ has the form
\beq
F_s(v,1) = \fr{1}{2\pi} F_S^B(\beta) \fr{df^{(0)}}{dv}\,\,.  
\eeq
Putting all these together yields
\beqa
\nn
\fr{1}{\pi}\fr{d\tl{f}^{(1)}}{dvdw} &=& F_h(v,w) +
\fr{F_s(v,w) -  \fr{1}{2\pi} F_S^B(\beta) \fr{df^{(0)}}{dv} }
{1-w} \\
\nn
& & + \fr{F_S^B(\beta)}{2\pi} \fr{df^{(0)}}{dvdw}
\lf[\ln(1-w_1)+\ln\lf(\fr{4v}{1-\beta^2}\rt)\rt] \\
&&+ \fr{F_\dl^B(\beta)}{2\pi} \fr{df^{(0)}}{dvdw}
+ F_\dl^{NB}(v) \dl(1-w)\,\,.
\eeqa
Writing
\beq
\fr{d\tl{f}^{(1)}}{dvdw} = \lf(\fr{df^{(1)}}{dvdw}\rt)_\dl +
\lf(\fr{df^{(1)}}{dvdw}\rt)_{N\dl},
\eeq
where the subscript $\dl$ denotes the part proportional to $\dl(1-w)$,
we have
\beqa
\nn
\fr{1}{\pi}\lf(\fr{df^{(1)}}{dvdw}\rt)_\dl &=&
 \fr{F_S^B(\beta)}{2\pi} \fr{df^{(0)}}{dvdw}
\lf[\ln(1-w_1)+\ln\lf(\fr{4v}{1-\beta^2}\rt)\rt] \\
&& + \fr{F_\dl^B(\beta)}{2\pi} \fr{df^{(0)}}{dvdw}
+ F_\dl^{NB}(v) \dl(1-w)
\eeqa
and
\beq
\fr{1}{\pi}\lf(\fr{df^{(1)}}{dvdw}\rt)_{N\dl} = F_h(v,w) +
\fr{F_s(v,w) -  \fr{1}{2\pi} F_S^B(\beta) \fr{df^{(0)}}{dv} }
{1-w}.
\eeq
The two simple $F$'s are given by
\beqa
F_S^B(\beta) &=& -4 \lf[ \fr{1+\beta^2}{2\beta}\ln x +1 \rt], \\
\nn
F_\dl^B(\beta) &=& - \fr{1+\beta^2}{2\beta} \Biggl\{ 2\ln x
-2 \lf[ \Lit\lf(\fr{-4\beta}{(1-\beta)^2}\rt) + \ln^2 x \rt] \Biggr. \\
\nn
& & \Biggl. + 2 \lf[-2\ln x \ln \beta + 2\Lit(-x) - 2\Lit(x) -
 \fr{\pi^2}{2}\rt]\Biggr\} 
\\
& & +2.
\eeqa
The other three $F$'s are somewhat lengthy and will not be presented
here as they can be directly inferred from the expressions given in
\cite{heavy}. $F_\dl^{NB}$ is the contribution from the virtual diagrams
which is not proportional to the Born term. $F_h(v,w)$ is proportional to
the first bracketed term in Eq.\ (30) of \cite{heavy} and $F_S(v,w)$ is
proportional to the second bracketed term of that same equation.
Both arise from gluon/photon bremsstrahlung.

It is standard \cite{Been} to divide the cross section (i.e.\ $f^{(1)}$) 
into two parts. Firstly, there is the virtual plus soft part,
\beq
\fr{df^{(1)}_{\rm V+S}}{dvdw} = \fr{df^{(1)}_{\rm V}}{dvdw}
+ \fr{df^{(1)}_{\rm S}}{dvdw},
\eeq
where $df^{(1)}_{\rm V}/dvdw$ denotes the virtual contribution and
$df^{(1)}_{\rm S}/dvdw$ is obtained by integrating the bremsstrahlung
contribution to $df^{(1)}/dvdw$ over the region
\beq
1 \geq w \geq w_{\rm 1, soft} \equiv 1 - \fr{m^2 \dl}{sv},
\eeq
then multiplying by $\dl(1-w)$.
We follow the definition of  the soft
parameter, $\dl$, given in \cite{Been} such that the gluon/photon radiated
becomes arbitrarily soft by making $\dl$ arbitrarily small. 
Since this takes into account all virtual corrections and soft radiation,
the hard radiation may be taken into account by integrating
$df^{(1)}/dvdw$ in the region $w_1 \leq w \leq w_{\rm 1, soft}$.
Since we never reach $w=1$, $F_s(v,w)/(1-w)_+ = F_s(v,w)/(1-w)$ in the
hard radiation integration. We thus define $df^{(1)}_{\rm H}/dvdw$
as being $df^{(1)}/dvdw$ in the region $w_1 \leq w \leq w_{\rm 1, soft}$,
where the $\dl(1-w)$ terms do not contribute:
\beq
\fr{df^{(1)}_{\rm H}}{dvdw} = \fr{df^{(1)}}{dvdw}
(w_1 \leq w \leq w_{\rm 1, soft}).
\eeq
It is not necessary to define $df^{(1)}_{\rm H}/dvdw$ outside that 
region since it is never evaluated there.

The sum of the (integrated over some region) hard and soft contributions 
so defined is independent of $\dl$ in the limit $\dl\RA 0$ and this
method of separation is referred to as the {\em phase space slicing
method}.
As one might expect, there is a close 
relation between $df^{(1)}_{\rm V+S}/dvdw$ and 
$(df^{(1)}/dvdw)_\dl$ as well as between $df^{(1)}_{\rm H}/dvdw$ and
$(df^{(1)}/dvdw)_{N\dl}$. We now give explicitly the 
necessary conversion terms.

It is straightforward to show that $d\sg_{\rm S}/dvdw$ is obtained from 
the term proportional to
$d\sg_{\rm LO}/dvdw$ in Eq.\ (30)
of \cite{heavy} by making the substitution
\beq
\label{ea26}
\ln\lf(\fr{sv}{m^2}\rt) \RA \ln\dl.
\eeq
From this, we infer that the conversion term necessary to transform
our result into $f^{(1)}_{\rm V+S}$  is
\beqa
\label{dconv}
\nn
\fr{1}{\pi}\fr{df^{(1)}_{\rm S, conv}}{dvdw}
= \fr{1}{2\pi} \fr{df^{(0)}}{dvdw} F_S^B(\beta) 
& \biggl[ & \ln\dl - \ln\lf( \fr{4v}{1-\beta^2}\rt) \\
 &-& \ln(1-w_1) \biggr].
\eeqa
The transformation is simply
\beq
\label{ea28}
\fr{1}{\pi}\fr{df^{(1)}_{\rm V+S}}{dvdw} = 
\fr{1}{\pi}\lf(\fr{df^{(1)}}{dvdw}\rt)_\dl + 
\fr{1}{\pi}\fr{df^{(1)}_{\rm S, conv}}{dvdw}
\eeq
To convert from $(df^{(1)}/dvdw)_{N\dl}$ to $df^{(1)}_{\rm H}/dvdw$
we must add the following conversion term,
\beqa
\label{hardconvv}
\fr{1}{\pi}\fr{df^{(1)}_{\rm H}}{dvdw} &=& 
\fr{1}{\pi}\lf(\fr{df^{(1)}}{dvdw}\rt)_{N\dl} + 
\fr{1}{\pi}\fr{df^{(1)}_{\rm H, conv}}{dvdw} \\ 
\nn
 &=& \fr{1}{\pi}\lf(\fr{df^{(1)}}{dvdw}\rt)_{N\dl} + 
\fr{1}{2\pi} \fr{F_S^B(\beta) df^{(0)}/dv }{1-w}.
\eeqa
Note that both soft and hard conversion terms are correctly determined
by taking $d\ha{f}^{(1)}/dvdw$ into account. 
Using (\ref{ea26}), we reproduced the result of \cite{Been} for 
$df^{(1)}_{\rm S, unp}/dvdw$ and using (\ref{ea28}), we 
reproduced the result of \cite{Been} for $df^{(1),{\rm unp}}_{\rm V+S}/dvdw$.
Further checks on $df^{(1)}/dvdw$ will be discussed in the next section. 

The variables $v,w$ are suitable for performing analytical integration
of the cross section (at least for the single integral part). They are
not, on the other hand, suitable for performing series expansions of
the integrated cross section about $\beta=0$. The reason is that,
in these variables, the integration limits depend on $\beta$ so that
the series expansion of the integrated cross section does not follow
straightforwardly from the series expansion of the differential cross
section, and we only have complete analytical results for the differential
cross sections. Otherwise, we could just expand the final integrated result.
The above will become clear in the following sections.
This approach also allows for cross checking; when one first expands the 
differential cross section and then integrates, the result should coincide
with that obtained by directly expanding the analytically integrated 
cross section. We will check this requirement for the single integral
part, for which we do have analytical results.

At this point, we introduce a new set of variables, $\tau$ and $\omg$,
suitable for performing series expansions in $\beta$.
They are defined through
\beq
\label{e28}
v\equiv \fr{1}{2}(1+\tau\beta), \SSPP w\equiv 1-c_\beta(\tau)\beta^2
(1-\omg),
\eeq
where
\beq
c_\beta(\tau) \equiv \fr{1-\tau^2}{1-\tau^2\beta^2}.
\eeq
We note
\beq
\label{e30}
\fr{df^{(i)}}{d\tau d\omg} =
\fr{\beta^3 c_\beta(\tau)}{2} \fr{df^{(i)}}{dvdw}.
\eeq
Because of the factor $c_\beta(\tau)$, $df^{(i)}/d\tau d\omg$ will
never get a part proportional to $\dl(1\pm \tau)$.
Defining
\beq
c_u \equiv c_\beta(\tau) (1-\omg) (1+\tau \beta),
\eeq
the invariants are given by
\beqa
\nn t &=& -\fr{s}{2} (1-\tau\beta), \SSPP u = -\fr{s}{2} (1+\tau\beta
-\beta^2 c_u), \\ 
\nn s_2 &=&  \fr{s}{2} c_u \beta^2, \SSPP 
T = -\fr{s}{4} (1-2\tau\beta+\beta^2), \\ 
\nn
U &=&  -\fr{s}{4} [1+2\tau\beta + \beta^2 (1-2c_u)], \\
S_2 &=& \fr{s}{4} [1+\beta^2(2c_u-1)].
\eeqa
In terms of $\tau$ and $\omg$, the LO term has the form
\beqa
\nn
\fr{1}{2\pi}\fr{df^{(0)}(j)}{d\tau d\omg}
&=& \fr{\beta \dl(1-\omg)}{(1-\tau^2\beta^2)^2} [ 2(1-\beta^4)
- j (1+\tau^2\beta^2) \\
\label{ea35}
&& \times (1+\tau^2\beta^2-2\beta^2)] \\ \nn
&=& \dl(1-\omg) [(2-j)\beta + (2j+4\tau^2-4j\tau^2)\beta^3 \\
&& + {\cal O}(\beta^5)].
\eeqa
We see explicitly that the $j=0,1$ 
differential cross sections, in terms of these
variables, vanish by order $\beta$ in the limit $\beta\RA 0$, while the
$j=2$ differential cross section is order $\beta^3$.

The conversion term (\ref{dconv}) becomes
\beqa
\label{dconvp}
\nn
\fr{1}{\pi}\fr{df^{(1)}_{\rm S, conv}}{d\tau d\omg} =
\fr{1}{2\pi}\fr{df^{(0)}}{d\tau d\omg}F_S^B(\beta)
\biggl[ && \ln\dl - \ln\lf(\fr{2(1+\tau\beta)}{1-\beta^2}\rt) \\
&& - \ln[\beta^2 c_\beta(\tau)]
\biggr]
\eeqa
and the conversion term (\ref{hardconvv}) becomes
\beq
\label{hardconvt}
\fr{1}{\pi}\fr{df^{(1)}_{\rm H, conv}}{d\tau d\omg} =
\fr{1}{2\pi} \fr{F_S^B(\beta) df^{(0)}/d\tau }{1-\omg}.
\eeq

%\vglue 1cm
%\begin{center}\begin{large}\begin{bf}
\section{ANALYTIC INTEGRATION OF THE DELTA FUNCTION PART}
%\end{bf}\end{large}\end{center}
%\vglue .3cm

The only complete analytical results
for the differential cross sections were presented in \cite{heavy}. Analytical
results for the virtual+soft part were presented in \cite{Been} for the
unpolarized case and in \cite{Jik} for the polarized case
(where the virtual and soft parts are given separately, in terms of
various functions).
Still, not even the virtual+soft part has previously been integrated
(over fermion angle) analytically. In this section, we present such an
analytical integration. We were not able to integrate the non delta function
(or hard) part analytically, in a straightforward fashion, and reserve that
for future work. 

The integrated cross section (or $f^{(i)}$) is obtained via
\beq
\label{ifexp}
f^{(i)} = \int_{v_1}^{v_2}\!\! dv \int_{w_1}^1 \!\! dw \fr{df^{(i)}}{dvdw}
= \int_{-1}^{1} \!\! d\tau \int_0^1 \!\! d \omg 
\fr{df^{(i)}}{d\tau d\omg}\,\, ,
\eeq
where
\beq
v_1 = \fr{1}{2} (1-\beta), \SSP v_2 = \fr{1}{2} (1+\beta).
\eeq
Let $\theta_3$ be the angle between $p_3$ and $p_1$ in the 
$\gm\gm$ c.m. Then $\theta_3$ is given by
\beqa
\cos\theta_3 &=& - \fr{1-v-vw}{\sqrt{(1-v+vw)^2+\beta^2-1}} \\
&=& \fr{2\tau-\beta c_u}{\sqrt{4-c_u(4-\beta^2 c_u)}}\,\,.
\eeqa 
Thus,  
\beq
\label{ea40}
\cos\theta_3 = - \fr{1-2v}{\beta} =  \tau, \SSPP \text{ for } w=\omg=1.
\eeq
We see that for $\beta\RA 0$, $\cos\theta_3$ varies rapidly with
$v$, while it is simply equal to $\tau$. This is why the phase space
in $v$ becomes vanishingly small by order $\beta$. Similarly, from
(\ref{ifexp}) and (\ref{e19}), or (\ref{e28}),
we see that the $w$ phase space is order $\beta^2$.
Thus, the double integration over $v$ and $w$ is order $\beta^3$, in accord
with (\ref{e30}).

The integration of (\ref{eb16}) or (\ref{ea35})
is rather straightforward, yielding the LO term,
\beq
\fr{f^{(0)}(j)}{2\pi} = 2\beta(1+\beta^2) - 6\beta j - (1-\beta^4
+ 2j) \ln x.
\eeq
Since
\beq
\ln x = -2 \sum_{k=0}^{\infty} \fr{\beta^{2k+1}}{2k+1}
= -2\beta -2\fr{\beta^3}{3} - \cdots,
\eeq
we have
\beqa
\fr{f^{(0)}(j)}{2\pi} &=&
2(2-j)\beta + \fr{4(2+j)}{3}\beta^3 \\ \nn
&& + 2 \sum_{k=2}^{\infty}
\lf( \fr{-1}{2k-3} + \fr{1+2j}{2k+1}   \rt) \beta^{2k+1},
\eeqa
so that $f^{(0)}(0,1)$ are order $\beta$ and $f^{(0)}(2)$ is order
$\beta^3$. Also, we see that only $f^{(0)}(0)$ is finite in the limit
$\beta\RA 1$ and it approaches
\beq
\label{ef47}
f^{(0)}(0)\RA 8\pi, \SSP \text{ for } \beta\RA 1\,\,.
\eeq
This is because the $1-\beta^4$ term in (\ref{ea35}) keeps the $j=0$
channel finite. For $j=2$, the cross section vanishes for {\em exactly}
$\tau=\pm 1$, as required by angular momentum conservation along the
$\gm\gm$ axis, but for $\beta\RA 1$ the part proportional to $j$ goes 
like $(1+\tau^2)/(1-\tau^2)$ as soon as we move away from {\em exactly}
$\tau=\pm 1$ and is hence not integrably finite at $\beta=1$. 
In order that the $j=0$ cross section
be nonvanishing for $\tau=\pm 1$, where its maximum lies,
 the $f$ and $\B{f}$ must have opposite spins
by angular momentum conservation, leading to $m^2/s$ ($\sim 1-\beta^2$)
suppression in the numerator. 
The fact that the LO $j=0$ cross section continues to be $1-\beta^2$
suppressed for $\tau\neq \pm 1$ follows from symmetry arguments
\cite{Bord}. This exactly 
compensates the $t$-channel singularity in the propagator, 
leading to a finite $f^{(0)}(0)$
for $\beta\RA 1$. 
Of course, for $\tau\neq\pm 1$, the $j=0$ differential
cross section will vanish like $(1-\beta^2)/(1-\tau^2)^2$, making it 
unobservable in LO, for $\beta\RA 1$.
So,  had we taken the limit $\beta\RA 1$ from the 
beginning, the $j=0$  cross section would have vanished 
identically. 
Hence the nonzero $f^{(0)}(0)$ in the $\beta\RA 1$ limit
is a remnant of using the fermion mass as a ``regulator''.

Near threshold, the $1-\beta^2$ suppression of the $j=0$ channel
will not be significant, hence the major constraint will come from
angular momentum conservation in the forward and backward directions
which will lead to suppression of the $j=2$ cross section there. 
The $j=0$ cross section reaches its maximum in those configurations,
however. 
Thus, we can clearly understand the feature of the numerical results for 
top quark production in \cite{heavy} which show that imposing angular cuts
in the direction of the beam pipe has a greater effect on the $j=0$
channel than on the $j=2$ channel.

We denote the single and double integral contributions to $f^{(1)}$
by
\beqa
\label{ea42}
f^{(1)}_{si/di} &\equiv& \int_{v_1}^{v_2} \,\, dv \int_{w_1}^1 \,\,
dw  \lf( \fr{df^{(1)}}{dvdw} \rt)_{\dl/N\dl}
\\ 
\label{ea42p}
&=& \int_{-1}^{1} \,\, d\tau \int_0^1 \,\,
d\omg  \lf( \fr{df^{(1)}}{d\tau d\omg} \rt)_{\dl/N\dl}.
\eeqa
We performed the single integration using (\ref{ea42}), as opposed to 
(\ref{ea42p}). It turned out to be quite lengthy and involved. We did
not check to see whether using (\ref{ea42p}) simplifies the calculation.
This question is probably more relevant to the double integration, however.
Our final (simplified) result is 
\beqa
\nn
\fr{f_{si}^1}{\pi} &=&
 a_1 \pi^2/6 + a_2 \Lit(x) + a_3 \Lit(-x) + \fr{\ln(x)}{\beta} 
 [a_4 \Lit(x) \\
\nn &&  + a_5 \Lit(-x)]  + a_6\ln[(1 + \beta)/2]  [\pi^2/6 
+ 2 \Lit(-x)] \\
\nn && + a_7 \ln(x) \pi^2/6 
 + a_8 \{ - 3 \ln^2[(3 + \beta^2)/4]  \ln(x) \\
\nn && + 5 \ln[(1 - \beta^2)/4] \ln[(3 + \beta^2)/4] \ln(x) 
\\ \nn
&& + 2 \lih[-2 x/(1 + \beta)] - 2 \lih[-2/(x (1 - \beta))] 
\\ \nn
&& + 2 \Lit[(1 - \beta)^2/(3 + \beta^2)] \ln[2/(x (1 - \beta))]  
\\ \nn
&& - 2 \Lit[(1 + \beta)^2/(3 + \beta^2)] \ln[2 x/(1 + \beta)] \}
\\ 
\label{ea43}
&& +  a_9 \{\lih[(1 + \beta)/2] - \lih[(1 - \beta)/2] \} 
\\ \nn
&& + a_{10} \ln(x) + a_{11} \ln^3(x)/\beta + a_{12} \ln^2(x)/\beta \\
\nn &&  + a_{13} \ln^2[(1 + \beta)/2] \ln(x) 
+ a_{14} \beta \\
\nn && + [a_{15} + a_{16} \ln^2(x)/\beta^2] \beta \ln(\beta) \\
\nn &&
+ [a_{17} \beta +   
a_{18} \ln(x) + a_{19} \ln^2(x)/\beta] \ln[(1 + \beta)/2].
\eeqa
The coefficients $a_i(j)$ are given in Appendix A and are rational
polynomials in $\beta$, finite as $\beta\RA 0$. Terms of the form
$F(\beta)-F(-\beta)$ will vanish for $\beta=0$. Noting that
\beq
\Lit(1)=\fr{\pi^2}{6}, \SSP \Lit(-1)=-\fr{\pi^2}{12},
\eeq
we see that only the terms proportional to $a_1$, \ldots, $a_5$ may
contribute at threshold, which is the case for $j=0$. 
Indeed, one finds the correct threshold correction from those 
terms alone. The relevant series
expansions will be given in the next section. As we shall see in 
Section V, the double integral series expansion starts at order 
$\beta^3$.

Two independent determinations of (\ref{ea43}) were made using 
Mathematica \cite{Math} and REDUCE \cite{REDUCE}. 
That software could not evaluate
certain integrals which can be found in \cite{Lewin}. 
It was verified that
the analytically integrated result agreed numerically with the numerically
integrated result. In the next section, we will show how one can use the
series expansion as a very solid check as well.

Perhaps the most convincing check of (\ref{ea43}) and the analytical result
for $df^{(1)}/dvdw$ (or $d\sg_{\rm NLO}/dvdw$) obtained in \cite{heavy} is the
excellent numerical agreement with tabulated results for $f^{(1)}$
existing in the literature. The only existing analytical results, aside from
those in \cite{heavy}, are the expressions for $df^{(1)}_{\rm V+S}/dvdw$
(i.e.\ $(df^{(1)}/dvdw)_{\dl}$), $df^{(1)}_{\rm S}/dvdw$  
given in \cite{Been} for the unpolarized case (using dimensional 
regularization), with which we agree exactly,
and similar expressions for the
polarized case in \cite{Jik} 
(obtained using a gluon energy cut and a small gluon
mass as infrared regulator). The latter are not quite in a form suitable
for direct analytical comparison.
There have been no other
analytical results presented for $(df^{(1)}/dvdw)_{N\dl}$ in the 
polarized or unpolarized cases. Hence we must perform the above mentioned
numerical checks.

Define
\beq
z\equiv \fr{\sqrt{s}}{2m} = \fr{1}{\sqrt{1-\beta^2}} \longleftrightarrow
\beta = \sqrt{1-1/z^2}\,\, .
\eeq
In Table \ref{TabI} we give numerically 
computed values for $f^{(1)}_{\rm unp}$,
$f^{(1)}_{\rm pol}$, $f^{(1)}(+,+)$, $f^{(1)}(+,-)$ as well as the
specific contributions from all the $f^{(1)}_{si}$ and $f^{(1)}_{di}$ to the
corresponding $f^{(1)}$, for various values of $1.2 \leq z \leq 20$. 
The result
at $z=1$ is given exactly by the series expansions presented in the next
section. We also indicate the number of significant figures, n.s., following
the decimal point, in $f^{(1)}_{di}$ (and $f^{(1)}$). 

We find it 
useful to describe how the values in Table \ref{TabI} were obtained in
order that one may see clearly which numbers have been rounded and how. 
The values for $f^{(1)}_{si,{\rm unp}}$, $f^{(1)}_{si,{\rm pol}}$
were obtained using (\ref{ea43}) which permits arbitrary precision, using
a package like Mathematica. The values of $f_{si}^{(1)}(+,+)$ and
$f_{si}^{(1)}(+,-)$ were obtained adding/subtracting the values of
$f^{(1)}_{si,{\rm unp}}$, $f^{(1)}_{si,{\rm pol}}$ so obtained.
The $f^{(1)}_{di,{\rm unp}}$, $f^{(1)}_{di,{\rm pol}}$ were obtained 
by numerical integration using
(\ref{ea42}) for the $N\dl$ part. From these, 
$f_{di}^{(1)}(+,+)$ and
$f_{di}^{(1)}(+,-)$ were obtained by 
adding/subtracting. Finally, $f^{(1)}(+,+)$,
$f^{(1)}(+,-)$ were obtained by adding the corresponding $f_{si}^{(1)}$,
$f_{di}^{(1)}$ (rather than adding/subtracting  $f^{(1)}_{\rm unp}$,
$f^{(1)}_{\rm pol}$); similarly for $f^{(1)}_{\rm unp}$,
$f^{(1)}_{\rm pol}$.
We did not check to see if (\ref{ea42p}) leads to
any reduction in computational time, for the precision obtained. The general
trend is that one needs more integration points as one goes to higher $z$.

The next issue is, of course, how well these values compare with other
tabulated values for $f^{(1)}$. Two other such tables exist at present.
The original one of \cite{Kuhn} gave $f^{(1)}_{\rm unp}$ for 
$z=2,3,4,5,10$; the value at $z=1$ being numerically
equal to the known threshold result, as given in the next section. Their
numerical values were obtained using the $f^{(1),{\rm unp}}_{\rm V+S}$
given in \cite{Been}, added numerically to $f^{(1)}_{\rm H, unp}$, 
determined there using the same methodology as \cite{Been}, which is 
equivalent
to our method. We find numerical agreement with \cite{Kuhn} 
to within the precision of those values, which is roughly at the order of
one part in $10,000$ or better. This can only be achieved with 
correct analytical results. Our calculation of  $f^{(1)}_{\rm pol}$ 
is identical in method (same integrals and structure)
to that of $f^{(1)}_{\rm unp}$ (at the differential
and integrated level), the only difference arising from different traces
due to the contraction with a polarized photonic tensor rather than
an unpolarized one. As two independent determinations of these traces were
performed, there is little room for any error in $f^{(1)}_{\rm pol}$.
Fortunately, we may directly check this assertion since the values 
of $f^{(1)}(+,+)$
and $f^{(1)}(+,-)$ for $z=2,3,4,5,10,20,50$ were tabulated 
in \cite{Jik}. There,
Monte Carlo methods were used, leading to accuracy at the level of better than
1\% in regions where the $f^{(1)}$ are sizable, but apparently not better than
$\pm 0.2$ or so in absolute error. This absolute error is noticeable only for
$f^{(1)}(+,+)$ and only for $z=2,3$, where $f^{(1)}(+,+)$ is small. 
To within the above accuracy, we are
in good agreement with \cite{Jik}. Since 
$f^{(1)}(+,-) =$ $f^{(1)}_{\rm unp} -$ $f^{(1)}_{\rm pol}$ and since we have
precision agreement with \cite{Kuhn} for $f^{(1)}_{\rm unp}$ and 
with \cite{Jik}
for $f^{(1)}(+,-)$, we conclude that our analytical results for
$df^{(1)}_{\rm pol}/dvdw$ of \cite{heavy} have been verified. In light of the
above, Table \ref{TabI} is seen to be the most complete and precise such
table at present.

We may convert from (\ref{ea43}) to $f^{(1)}_{\rm V+S}$ by adding the
following conversion term,
\beqa
\label{ef53}
\lefteqn{
\fr{f^{(1)}_{\rm V+S}}{\pi} = \fr{f^{(1)}_{si}}{\pi} + 
\fr{f^{(1)}_{\rm S, conv}}{\pi}
} \\
\nn &=&\fr{f^{(1)}_{si}}{\pi} + 
 F_S^B(\beta)\Biggl\{ \lf[\ln\lf(\fr{1-\beta^2}{4}
\rt) + \ln \dl \rt] \fr{f^{(0)}}{2\pi} \\
\nn && +\beta(1+\beta^2-4j) +2\beta(1+\beta^2-3j)\ln\lf(
\fr{1+\beta}{2\beta^2}\rt) \\
\nn && + \lf[-\fr{3}{2} + \beta - 2\beta^2 
+ \beta^3 - \fr{\beta^4}{2} +j(4-3\beta+\beta^2)\rt]
\ln x \\
\nn && + \biggl[ \fr{1}{4} \ln^2 x + 2 \ln x \ln\lf(\fr{1+\beta}{2}\rt)
- 2 \Lit (x) - \Lit (-x)  \\
\nn && + \fr{\pi^2}{4}\biggr]
(1-\beta^4+2j)\Biggr\},
\eeqa
where $f^{(1)}_{\rm S, conv}$ follows from integrating 
$df^{(1)}_{\rm S, conv}/dv$,
given in (\ref{dconv}), or from integrating $df^{(1)}_{\rm S, conv}/d\tau$,
given in (\ref{dconvp}).

%\vglue 1cm
%\begin{center}\begin{large}\begin{bf}
\section{SERIES EXPANSION OF THE DELTA FUNCTION PART}
%\end{bf}\end{large}\end{center}
%\vglue .3cm

Besides providing a useful check of the analytical integration of the 
previous section, there are many reasons why it is useful and instructive
to series expand the differential and integrated cross sections about
$\beta=0$. In the absence of complete analytically integrated results, only
a series expansion about $\beta=0$ can be used to make (very) high precision
predictions in the $\beta\simeq 0$ region. One also sees the structure of
the cross section in a way that cannot be inferred from the 
non-expanded analytical results, which are somewhat complicated. 
From a practical viewpoint, having
``simple'' series expansions for the differential cross sections allows one
to do complete numerical studies in the region not too far above threshold
rather easily. This is because the resulting expansions only involve simple
polynomials and simple logarithms. We will address the issue of the region
of validity of the expansions as well.

The other issue is that of resummation. There are large correction 
terms at threshold which can be resummed. Having a 
series expansion of high enough order to be of practical use allows
one to explicitly perform resummations up to some order in $\beta$
while leaving the higher order 
terms the same. The net result would
be an equally simple series, improved via resummation so as to allow
one to go closer to threshold. This is beyond the scope of 
\newpage \noindent
this paper
as are other very near threshold effects. Suffice it to say that having
the threshold series expansion will facilitate these studies for those
interested.

Throughout, we will expand up to order $\beta^{10}$ (including 
$\beta^{11}\ln\beta$ terms). The expansion which exists in the 
literature (see \cite{Kuhn}) is only for $f^{(1)}_{\rm unp}$ and only goes
to order $\beta$. Going to order $\beta^{10}$ may seem excessive at
first, but we found it to be a good stopping point for several reasons.
Considerable structure arises beyond order $\beta$ which allows us to see
the general, all-orders in $\beta$, structure of the various series.
Also, one gains little in terms of precision by going to even higher
orders in $\beta$, without including several more terms. Then, the
series would start to become lengthy and cumbersome, reducing the
advantage over the analytical result in terms of ease of use. For certain
series, going much beyond $\beta^{10}$ would take a very large amount
of computer memory and run-time, not justifying the extra effort, as going
to order $\beta^{10}$ was a considerable task in itself. Finally, by going
to such a high order, we may stringently check the analytically integrated
single integral result of the previous section as will be described 
below.

We find that $d\tl{f}^{(1)}/d\tau d\omg$ may be expanded in the general form
\beq
\label{ea45}
\fr{d\tl{f}^{(1)}}{d\tau d\omg} = \sum_{i=0}^{\infty} \sum_{j=0}^1
c_{ij}(\tau,\omg) \beta^i \ln^j \beta.
\eeq
Therefore $f^{(1)}$ may be expanded as
\beq
\label{ea46}
f^{(1)} = \tl{f}^{(1)} = 
\sum_{i=0}^{\infty} \sum_{j=0}^1 d_{ij}
\beta^i \ln^j \beta,
\eeq
where the $d_{ij}$ are given by
\beq
 d_{ij} = \int_{-1}^{1} d\tau \int_0^1 d\omg \,  c_{ij}
(\tau,\omg).
\eeq
With the variables $v$ and $w$, the integration
limits depend on $\beta$, hence the above arguments do not hold.
So, one sees clearly the necessity of the change of variables. 

We convert from $(df^{(1)}/dvdw)_\dl$
to $(df^{(1)}/d\tau d\omg)_\dl$ using (\ref{e30}), which
modifies the overall factor via  $\dl(1-w) \RA
\beta^3 c_\beta(\tau) \dl(1-w)/2 =  \beta \dl(1-\omg)/2$.
Then, the results for the series expansions of $(df^{(1)}/d\tau d\omg)_\dl$
are, for $j=0$, 
\onecolumn \noindent
%\widetext
\beqa \label{eb57}
\nn
 \lefteqn{\fr{1}{\pi}\lf(\fr{df^{(1)}(+,+)}{d\tau d\omg}\rt)_{\dl} =} 
    \\ \nn
     & & \dl(1-\omg) \biggl\{
    2 \pi^2 + \beta (-20 +  \pi^2) + 2 \beta^2 \pi^2 (1 + 2 \tau^2)
    +  \beta^3 \Bigl(-9 \pi^2 (2 - 7 \tau^2) - 4 (22 + 45 \tau^2) 
    \\ \nn   &+& 
      96 \{(1 - 9 \tau^2) \ln(2) + \ln[4 \beta^2 (1 -  \tau^2)]\}\Bigr)/9 
    +  2 \beta^4 [-3 \pi^2 + 16 \tau + 3 \pi^2 \tau^2 (2 + 3 \tau^2)]/3
     \\ \nn
    &+&  \beta^5 \Bigl(225 \pi^2 (1 - 14 \tau^2 + 17 \tau^4) 
      + 4 [1394 - 25 (464 \tau^2 - 381 \tau^4)] 
      + 480 \{(-34 + 205 \tau^2 - 245 \tau^4) \ln(2) 
     \\ \nn &+& 
          2 (1 + 5 \tau^2) \ln[4 \beta^2 (1 -  \tau^2)]\}\Bigr)/225
     + 2 \beta^6 [16 \tau (6 + 35 \tau^2)/45 - \pi^2 (1 +  \tau^2) (1 
        +  \tau^2 - 4 \tau^4)]
     \\ \nn
    &+&  \beta^7 \Bigl[\pi^2 \tau^2 (7 - 34 \tau^2 + 31 \tau^4)
      + 
    4 \{-72244 + 49 \tau^2 [35068 - 5 \tau^2 (18680 - 12373 \tau^2)]\}/11025 
      \\ \nn
   &+& 
    32 \Bigl(\{227 + 7 \tau^2 [-473 +  \tau^2 (1315 - 973 \tau^2)]\} \ln(2) 
     + 
          [-26 + 7 \tau^2 (4 + 15 \tau^2)] 
 \\ \nn
  &\times &\ln[4 \beta^2 (1 
             -  \tau^2)]\Bigr)/105\Bigr] 
    + 2 \beta^8 \{ \pi^2 \tau^2 (-1 -  \tau^2) (2 +  \tau^2 - 5 \tau^4) 
     \\ \nn &+& 
     32 \tau [-39 + 7 \tau^2 (7 + 29 \tau^2)]/315\} 
    +  \beta^9  \Bigl[315 \pi^2 \tau^4 [17 +  \tau^2 (-62 + 49 \tau^2)] 
     \\ \nn &+& 
      4 \Bigl(830486 + 9 \tau^2 \{-2777328 + 
            7 \tau^2 [2146844 + 65 \tau^2 (-55762 + 29125 \tau^2)]\}
      \Bigr)/315 \\ 
    \nn &+& 
      32 \Bigl(-376 + 3 \tau^2 \{4899 +  \tau^2 [-26579 +  \tau^2 (46207
           - 25243 \tau^2)]\}\Bigr) 
       \ln(2) 
           + 64 \{-11 + 3 \tau^2 [-26 \\ \nn
    &+&  7 \tau^2 (3 + 10 \tau^2)]\} 
       \ln[4 \beta^2 (1 -  \tau^2)]\Bigr]/315 
    + 2 \beta^{10} \Bigl(\pi^2 \tau^4 (-1 -  \tau^2)
         (3 +  \tau^2 - 6 \tau^4) 
     \\ \nn &+& 
     32 \tau \{-55 +  \tau^2 [-455 +  \tau^2 (406 + 1455 \tau^2)]\}/1575
      \Bigr)
     \\ 
    &+&  64 \beta^{11}\ln(2 \beta) \Bigl(-122 + 11 \tau^2 \{-44 + 3 \tau^2 
        [-78 + 7 \tau^2 (8 + 25 \tau^2)]\}\Bigr) 
           /3465
     \biggr\}
\eeqa
and, for $j=2$, 
\beqa \label{eb58}
\nn
%\narrowtext \noindent
 \lefteqn{\fr{1}{\pi}\lf(\fr{df^{(1)}(+,-)}{d\tau d\omg}\rt)_{\dl} =} 
     \\ \nn
  && 2 (1 -  \tau^2) \dl(1-\omg) \biggl\{
    2 \beta^2  \pi^2  - 16 \beta^3   
    + \beta^4  \pi^2  (1 + 5 \tau^2) \\ \nn
    &+& 4 \beta^5  \{14 - 159 \tau^2 - 6 (7 - 19 \tau^2) \ln(2) \\ \nn
    &+&
    24 \ln[4 \beta^2 (1 -  \tau^2)]\}/9  
    - \beta^6  \{-32 \tau/3 + \pi^2 [1 \\ \nn
     &-&  \tau^2 (3 + 8 \tau^2)]\}
     - 2 \beta^7   \Bigl(1102 - 21355 \tau^2 + 37995 \tau^4 
     \\ \nn
     &+& 120 \{(-19 + 273 \tau^2 - 382 \tau^4) \ln(2)  
     \\ \nn
      &+&   (1 - 25 \tau^2) \ln[4 \beta^2 (1 -  \tau^2)]\}\Bigr)/225 \\ \nn
    &-&  \beta^8  \tau  [16 (3 - 85 \tau^2)/45 + \pi^2 \tau (2 - 5 \tau^2 
         - 11 \tau^4)]
     \\ \nn
    &+& 2 \beta^9   \{129611 - 805 \tau^2 [3721 - 2 \tau^2 (6781 - 6092 
        \tau^2)] \\ \nn
    & + &
      840 [-219 +  \tau^2 (5172 - 18803 \tau^2 + 16134 \tau^4)] \ln(2) 
     \\ \nn &+& 
      3360 (1 + 70 \tau^4) \ln[4 \beta^2 (1 -  \tau^2)]\}/11025  \\ 
    &+&  \beta^{10}  \tau [\pi^2 \tau^3 (-3 + 7 \tau^2 + 14 \tau^4)
     \\ \nn  &+& 
     16 (12 - 7 \tau^2 + 1057 \tau^4)/315] \\ \nn
    &+&  32 \beta^{11} \ln(2 \beta) [13 + 51 \tau^2 + 21 \tau^4 
     (1 + 55 \tau^2)] /315
     \biggr\} .
\eeqa
%\narrowtext \noindent
The remaining terms are of order $\beta^{11}$. Here, and throughout, we
group terms proportional to $\ln(2\beta)$ rather than $\ln\beta$ as in
(\ref{ea45}), (\ref{ea46}) for purposes of compactness. Of course, the
choice is quite arbitrary. 

We notice that the cross section is isotropic up to order $\beta$; the
angular ($\tau$) dependence enters only at order $\beta^2$. The LO term,
on the other hand, was isotropic up to order $\beta^2$. We also see that
the step function threshold behaviour 
\twocolumn \noindent
arises entirely from the $j=0$
channel, at the level of the differential cross section, since the $j=2$
channel starts at order $\beta^2$. From the $1-\tau^2$ overall factor,
and using (\ref{ea40}), we see that the delta function contribution
to the $j=2$ cross section vanishes at $\cos\theta_3=\pm 1$ as did
the LO cross section (\ref{ea35}). This vanishing is not obvious
from the exact analytical expressions, but simply reflects angular
momentum conservation along the $\gm\gm$ axis when $\omega=1$
($2\RA 2$ kinematics). The $j=0$ channel, on the other hand, becomes infinite
(but integrably finite) for $\cos\theta_3=\pm 1$ due to the 
$\ln(1-\tau^2)$ terms.

The expansions (\ref{eb57}), (\ref{eb58}) have rather simple structure
in that, aside from the $\ln(1-\tau^2)$ terms, the $c_{ij}(\tau)$ are 
simply polynomial in $\tau$. This amounts to considerable simplification
and reduction in computational time relative to the exact expressions,
especially after the non delta function part is added, where the 
simplification is even greater as we shall see in the next section. 
Adding the conversion term (\ref{dconvp}) gives $\fr{1}{\pi}
df^{(1)}_{\rm V+S}/d\tau d\omg$.

Two independent calculations of (\ref{eb57}), (\ref{eb58}) were performed
using Mathematica and REDUCE. The expansions were also checked 
numerically by subtracting them from the exact expressions. The difference
was checked to be of order $\beta^{11}$. This is most straightforwardly
done by taking rather small $\beta$.

Assuming we are working at $\beta$ where the series are sufficiently
accurate, one could easily analytically integrate (\ref{eb57}),
(\ref{eb58}) over a region of $\tau$ ($\cos\theta_3$) relevant to
some experiment, if desired, and implement angular
cuts analytically. After a suitable change of variables, the same
could be done for the hard radiation part, either analytically
or numerically.
Cuts on additional observables may be made by subtracting
off the unwanted configurations using the squared amplitudes given
in \cite{heavy} and Monte Carlo integration, for instance. Here, we simply
present the total integrated results. 

For the $j=0$ channel, we find
\beqa \label{ef59}
\nn
\lefteqn{\fr{1}{\pi}f^{(1)}_{si}(+,+) = } \\ 
\nn  &&  2 \biggl\{
      2 \pi^2 - (20-\pi^2) \beta + 10/3 \pi^2 \beta^2 + \beta^3/3 [-340/3 
        + \pi^2 \\ \nn
    && + 64 \ln(2 \beta)] + 8/15 \pi^2 \beta^4 + 4/15 \beta^5 
       [-6343/45 - \pi^2
     \\ \nn    
 && - 32 \ln(2) + 256/3 \ln(2 \beta)] - 104/105 \pi^2 \beta^6 \\ \nn
     && + 4/105 \beta^7 [-39163/315 - \pi^2 + 208/3 \ln(2 \beta)] \\ \nn
     &&- 88/315 
          \pi^2 \beta^8 
      + 4/9 \beta^9 \{ 1/7[-\pi^2/5 - 64/3 \ln(2)] \\ \nn
  && + 1/25[128 
         \ln(2 \beta)-43903/315] \}  
      - 488/3465 \pi^2 \beta^{10} \\ 
   && + 103232/51975 \beta^{11} \ln(2 \beta)
      \biggr\}
\eeqa
and for $j=2$,
\beqa \label{ef60}
\lefteqn{\fr{1}{\pi}f^{(1)}_{si}(+,-)=} \\
\nn   && 16/3 \biggl\{
      \pi^2 \beta^2 - 8 \beta^3 + \pi^2 \beta^4 + 32 \beta^5 [-289/720 
         + \ln(2)/5  \\ \nn  && + \ln(2 \beta)/3] 
       + \pi^2/7 \beta^6 + 6/5 \beta^7 [-3947/945 + 16/7 \ln(2) \\ \nn
         &&       + 32/9 \ln(2 \beta)] 
       + 29/105 \pi^2 \beta^8 + 4/15 \beta^9 [-823/45 + 8 \ln(2) \\ \nn
        &&   + 16 \ln(2 \beta)] 
       + 289/1155 \pi^2 \beta^{10} + 256/63 \beta^{11}  \ln(2 \beta)
      \biggr\}.
\eeqa
The results are indeed quite simple. We may obtain 
$\fr{1}{\pi}f^{(1)}_{\rm V+S}$ by adding the conversion term
(\ref{ef53}).

The strongest check comes from the fact that the expansions 
(\ref{ef59}), (\ref{ef60}) which come from integrating
(\ref{eb57}), (\ref{eb58}) agree exactly with the expression obtained
by expanding the analytically integrated result (\ref{ea43}) directly.
In this way, we simultaneously check all the above mentioned 
expressions, including our analytical integration (\ref{ea43}).
The expansions (\ref{ef59}), (\ref{ef60}) were also checked numerically
by subtracting them from the exact expression (\ref{ea43}) and
verifying that the difference was order $\beta^{11}$, as was done
for (\ref{eb57}), (\ref{eb58}).

%\vglue 1cm
%\begin{center}\begin{large}\begin{bf}
\section{SERIES EXPANSION OF THE NON DELTA FUNCTION PART}
%\end{bf}\end{large}\end{center}
%\vglue .3cm

Perhaps the most remarkable result of the series expansion is the
simplification of the non delta function part, whose original form
is the most lengthy part of the exact result, involving complicated
logarithms, etc\ldots Although the intermediate expressions were very
lengthy and considerable computational time was required, a large
degree of cancellation resulted in the following simple series. 
We convert from $(df^{(1)}/dvdw)_{N\dl}$
to $(df^{(1)}/d\tau d\omg)_{N\dl}$ by multiplying by $\beta^3 
c_\beta(\tau)/2$.
Then, for the
$j=0$ channel we find
\beqa \label{ef61}
\nn 
 \lefteqn{\fr{1}{\pi}\lf(\fr{df^{(1)}(+,+)}{d\tau d\omg}\rt)_{N\dl} =} 
    \\ \nn && - 8/3 (1- \tau^2) \biggl\{ 
    4 \beta^3 + 4 \tau \beta^4 -1/5 \beta^5 [27-43 \omg \biggr. \\
\nn &&
  - \tau^2(63-43 \omg)]
   - 2/5 \tau \beta^6 [15-43 \omg -  \tau^2(53-43 \omg)] \\
\nn &&  +1/35 \beta^7 [2( -87 -108 \omg + 179 \omg^2) -  \tau^2(1133-2353 
    \omg \\
\nn && + 716 \omg^2) 
            +  \tau^4(1499 - 2137 \omg + 358 \omg^2) ]
    \\ \nn
   && + 8/35 \tau \beta^8 [ 70 - 259 \omg + 199 \omg^2 - 2 \tau^2 (133 
       - 416 \omg  \\ 
 && + 199 \omg^2)
              +  \tau^4 (339 - 573 \omg + 199 \omg^2) ] \\
 \nn
   && - 2/105 \beta^9 [ 338 - 1363 \omg + 2222 \omg^2 - 1189 \omg^3
    \\ \nn
             && +  \tau^2 (-1639 + 11135 \omg - 13399 \omg^2 + 3567 \omg^3)
    \\ \nn
             && -  \tau^4 (-8663 + 27244 \omg - 20132 \omg^2 + 3567 \omg^3)
    \\ \nn
             && +  \tau^6 (-7606 + 17472 \omg - 8955 \omg^2 + 1189 \omg^3) ]
    \\ \nn
   && + 2/105 \tau \beta^{10} [ - 462 + 4317 \omg - 9396 \omg^2 + 5221 \omg^3
    \\ \nn
    && + 3 \tau^2 (2529 - 12399 \omg + 15371 \omg^2 - 5221 \omg^3)
    \\ \nn
    && + 3 \tau^4 ( - 6022 + 23659 \omg - 21346 \omg^2 + 5221 \omg^3)
    \\ \nn
    && \biggl. +  \tau^6 (13897 - 38097 \omg + 27321 \omg^2 - 5221 \omg^3)]
    \biggr\}
\eeqa
and, for $j=2$,
\beqa \label{ef62}
 \nn 
 \lefteqn{\fr{1}{\pi}\lf(\fr{df^{(1)}(+,-)}{d\tau d\omg}\rt)_{N\dl} =} 
 \\ \nn  && - 8/3 (1- \tau^2) \biggl\{ 
    4 \beta^5 [1+\omg +  \tau^2(1-\omg)] -  8 \tau \beta^6  (1-\omg) \biggr. 
\\
\nn && \times (1- 
      \tau^2)
    + 2/5 \beta^7 [ 3(1 - 6 \omg+ 7 \omg^2) - 2 \tau^2(14 - 31 \omg \\
\nn &&
    + 21 \omg^2) 
    +  \tau^4(73 - 44 \omg + 21 \omg^2) ]
    - 2/5 \tau (1- \tau^2) \beta^8 \\
\nn && \times [ -53 + 30 \omg - 35 \omg^2 
           +  \tau^2(169 - 154 \omg + 35 \omg^2) ] \\
\nn &&
   - 1/35 \beta^9 [ 152 - 226 \omg + 385 \omg^2 - 415 \omg^3 
    \\ \nn
           && +  \tau^2(-2688 + 3890 \omg - 3451 \omg^2 + 1245 \omg^3)
    \\ \nn
           && -  \tau^4(-7192 + 8782 \omg - 5747 \omg^2 + 1245 \omg^3)
    \\ \nn
           && +  \tau^6(-5792 + 5118 \omg - 2681 \omg^2 + 415 \omg^3) ]
    \\ \nn
    && + 2/35 (1- \tau^2) \tau \beta^{10} [ - 420 + 1035 \omg - 1475 \omg^2 
           + 904 \omg^3
    \\ \nn  
      && + 2 \tau^2 (2091 - 2569 \omg + 2215 \omg^2 - 904 \omg^3)
    \\ 
      && \biggl. +  \tau^4 ( - 6390 + 6971 \omg - 2955 \omg^2 + 904 \omg^3) ]
       \biggr\}.
\eeqa

We notice the absence of any logarithms, including powers of $\ln\beta$.
The structure is fairly predictable as well. We see that the series
begin at order $\beta^3$ and $\beta^5$ respectively, so that their 
effect will be negligible very near to threshold. On the other hand,
the large coefficients imply that they soon become noticeable for small
$\beta$. We may obtain 
$\fr{1}{\pi} df^{(1)}_{\rm H}/d\tau d\omg$ by adding to 
(\ref{ef61}), (\ref{ef62}) the conversion term (\ref{hardconvt}).

Two independent determinations of (\ref{ef61}), (\ref{ef62}) were
performed using Mathematica and REDUCE. These expressions were also
checked numerically analogously to the delta function part of the
differential cross section.

The integration of (\ref{ef61}), (\ref{ef62}) over $\tau$, $\omg$ is
straightforward and we obtain
\beq \label{ef63}
\fr{1}{\pi}   f^{(1)}_{di}(+,+) = \fr{-128}{9} \beta^3 -  
 \fr{448}{225} \beta^5 
    + \fr{34624}{2205} \beta^7 + \fr{42368}{3675} \beta^9,
\eeq 
\beq \label{ef64}
\fr{1}{\pi}   f^{(1)}_{di}(+,-) = \fr{-1024}{45} \beta^5 - 
\fr{2816}{525} \beta^7 
    - \fr{134656}{19845} \beta^9.
\eeq
These are remarkably simple results, which suggest that the exact 
integrated result for $f^{(1)}_{di}$ is not too complicated. We notice
the vanishing of the coefficients of the even powers of $\beta$. This
follows from the antisymmetry in $\tau$ of the corresponding terms in
the differential cross section. 
To convert to $f^{(1)}_{\rm H}$, we must integrate the conversion
term (\ref{hardconvv}) (or (\ref{hardconvt})), over $v$ and between
$w_1 \leq w \leq w_{\rm 1, soft}$. This yields
\beqa
\fr{f^{(1)}_{\rm H}}{\pi} &=& \fr{f^{(1)}_{di}}{\pi}
+ \fr{f^{(1)}_{\rm H, conv}}{\pi} \\
\nn 
 &=& \fr{f^{(1)}_{di}}{\pi}
- \fr{f^{(1)}_{\rm S, conv}}{\pi},
\eeqa
where $f^{(1)}_{\rm S, conv}$ is given in (\ref{ef53}).
This verifies the cancellation of the $\dl$ dependence 
of $f^{(1)}_{\rm H} + f^{(1)}_{\rm S}$    in the limit
$\dl\RA 0$. Implicitly we were working in this limit since we integrated
the non delta function part over all $\tau$, $\omg$.

We checked  (\ref{ef63}), (\ref{ef64}) against the numerically integrated
result and again found the difference was order $\beta^{11}$. In the 
next section we will tabulate the numerical errors on 
the series expansions for $f^{(1)}$
(total) for various values of $\beta$, relative to the numerical
result obtained from the exact expressions.

%\vglue 1cm
%\begin{center}\begin{large}\begin{bf}
\section{TOTAL SERIES RESULTS AND NUMERICAL PARAMETRIZATIONS}
%\end{bf}\end{large}\end{center}
%\vglue .3cm

We are now in a position to study the total cross section, by combining
the results of the previous sections. Adding (\ref{ef59}) and
(\ref{ef63}) gives the series for the $j=0$ total cross section
\beqa  \nn
    \lefteqn{\fr{1}{\pi} f^{(1)}(+,+)=} \\
\nn &&  2 \biggl\{
      2 \pi^2 - (20-\pi^2) \beta + 10/3 \pi^2 \beta^2 + \beta^3/3 [-404/3 
         + \pi^2 \biggr.  \\ \nn
    && + 
      64 \ln(2 \beta)] + 8/15 \pi^2 \beta^4 + 4/15 \beta^5 [-6511/45 - \pi^2
    \\ \nn
      && - 32 \ln(2) + 256/3 \ln(2 \beta)] - 104/105 \pi^2 \beta^6
     \\ \nn
      && + 4/105 \beta^7 [25757/315 - \pi^2 + 208/3 \ln(2 \beta)] 
          - 88/315 \pi^2 \beta^8
     \\ \nn
      && + 4/9 \beta^9 \{1/7[-\pi^2/5 - 64/3 \ln(2)] + 1/25[128 
             \ln(2 \beta) \\ \nn 
     && +407639/2205]\}
   - 488/3465 \pi^2 \beta^{10} \\ 
    && + 103232/51975 \beta^{11} \ln(2 \beta)
      \biggr\}
\eeqa
and adding (\ref{ef60}), (\ref{ef64}) gives the series for the $j=2$
total cross section
\beqa \nn
\lefteqn{\fr{1}{\pi} f^{(1)}(+,-) = } \\ 
\nn &&  16/3 \biggl\{
      \pi^2 \beta^2 - 8 \beta^3 + \pi^2 \beta^4 + 32 \beta^5 [-77/144 + 
          \ln(2)/5  \\ \nn 
    && + \ln(2 \beta)/3] 
      + \pi^2/7 \beta^6 + 6/5 \beta^7 [-677/135 + 16/7 \ln(2) \\ \nn
    && + 32/9 
            \ln(2 \beta)]
       + 29/105 \pi^2 \beta^8 + 4/15 \beta^9 [-16949/735  \\ \nn 
    && + 8 \ln(2) 
       + 16 \ln(2 \beta)]
       +  289/1155 \pi^2 \beta^{10} \\ 
  && + 256/63 \beta^{11} \ln(2 \beta)
      \biggr\}.
\eeqa
Such simple expressions indeed make numerical studies not too far above 
threshold rather straightforward. We can get an idea of how well these
series work for typical $\beta$ by comparing with numerically 
calculated values of $f^{(1)}$.

In Table \ref{TabII} we present the fractional error on the series
for $f^{(1)}(+,+)$, $f^{(1)}(+,-)$, $f^{(1)}_{\rm unp}$ relative to
the result obtained using numerical integration, for various values
of $z$ in the region $1.05 \leq z \leq 1.4$. For $z\lesssim 1.05$, the
series expansions are more accurate than the numerical results.
At $z=1.05$, the errors are at the $10^{-7}-10^{-6}$ level. For
$z=1.2$ they are at the $10^{-4}-10^{-3}$ level and for
$z=1.4$ they are at the $10^{-3}-10^{-2}$ level. The errors on
$f^{(1)}(+,+)$ are at the lower end, while the errors on 
$f^{(1)}(+,-)$ are at the higher end and those for 
$f^{(1)}_{\rm unp}$ lie in between. This is good because, as we shall
see in the next section, in determining $\alpha_s$ via top quark
production at a $\gm\gm$ collider, it is the $j=0$ and unpolarized
channels which are of interest, the $j=0$ channel being the most interesting
one. With precision of better than one percent for $z\leq 1.4$, we 
have sufficient accuracy to use the series expansions (differential
in particular) to perform easy numerical studies relevant to 
top quark production at a $\gm\gm$ collider of $\sqrt{s}\lesssim 500$ $\GeV$.
As we shall see, for the $\alpha_s$ determination, going to much higher 
energies is not useful since the determination is best done near
$z=1.2$ ($\sqrt{s}\simeq 420$ $\GeV$).

It is also useful to be able to parametrize $f^{(1)}$ to good 
accuracy for larger $\beta$, relevant for bottom and charm quark
production at intermediate energies or top quark production at very
high energies. This was done by fitting numerically computed values
of $f^{(1)}$.
We divide the parametrizations into 3 regions: a low energy region
($1\leq z\leq 1.5$ or $0\leq\beta\leq 0.7454$), 
an intermediate energy region ($1.5<z\leq 5$)
and a high energy region ($5<z\leq 20$). We will denote the corresponding 
$f^{(1)}$ as $f^{(1),le}$, $f^{(1),ie}$ and $f^{(1),he}$, respectively.

The various forms for the parametrizations are
\beqa \label{ef67}
\nn
f^{(1),le}(+,+) &=&  2\pi\biggl[2\pi^2-(20-\pi^2)\beta + \fr{10\pi^2}{3}\beta^2
\\ \nn 
&& +\fr{64}{3}\beta^3\ln\beta\biggr] 
 + \sum_{i=3}^7 c_i\beta^i, \\
\nn f^{(1),ie}(+,+) &=& \sum_{i=0}^6 c_i (z-1.5)^i, \\ 
f^{(1),he}(+,+) &=& \sum_{i=0}^4 c_i (z-5)^i
\eeqa
and 
\beqa \label{ef68}
\nn
f^{(1),le}(+,-) &=&  \fr{16\pi}{3}
\lf[\pi^2\beta^2 -8\beta^3
+\pi^2\beta^4\rt] 
 + \sum_{i=5}^{10} c_i\beta^i, \\
\nn f^{(1),ie}(+,-) &=& \sum_{i=0}^4 c_i (z-1.5)^i, \\
f^{(1),he}(+,-) &=& \sum_{i=0}^3 c_i (z-5)^i.
\eeqa 
The $c_i$ are given in Appendix B.
In the low energy region, where high accuracy is required, the
parametrizations are accurate to $\lesssim 0.01$\%, with the errors
being the largest near the higher end of the region. The leading
terms, given analytically, guarantee the correct threshold behaviour
as they are just those in the exact series expansion. 
As mentioned earlier in connection with the 
series expansions, one can explicitly perform resummations on those
terms. Thus one could modify the above parametrizations to include
resummation effects without changing the higher order coefficients.
Here, we simply present the one-loop corrections.

In the intermediate energy region, $f^{(1),ie}(+,-)$ is accurate to
$\lesssim 0.1$\%. $f^{(1),ie}(+,+)$ is accurate to $\lesssim 1$\%,
except very near $f^{(1),ie}(+,+)=0$, which occurs for $z\simeq2.15$,
$3.15$. There, the absolute errors remain small, but of course the
fractional error is larger. In the high energy region, $f^{(1),he}(+,-)$ 
is accurate to $\lesssim 0.05$\%, while  
$f^{(1),he}(+,+)$ is accurate to $\lesssim 0.5$\%. The above errors
are rather conservative and one can not distinguish the parametrizations
from the exact results for practical purposes.

\begin{figure}
\centerline{\epsfig{file=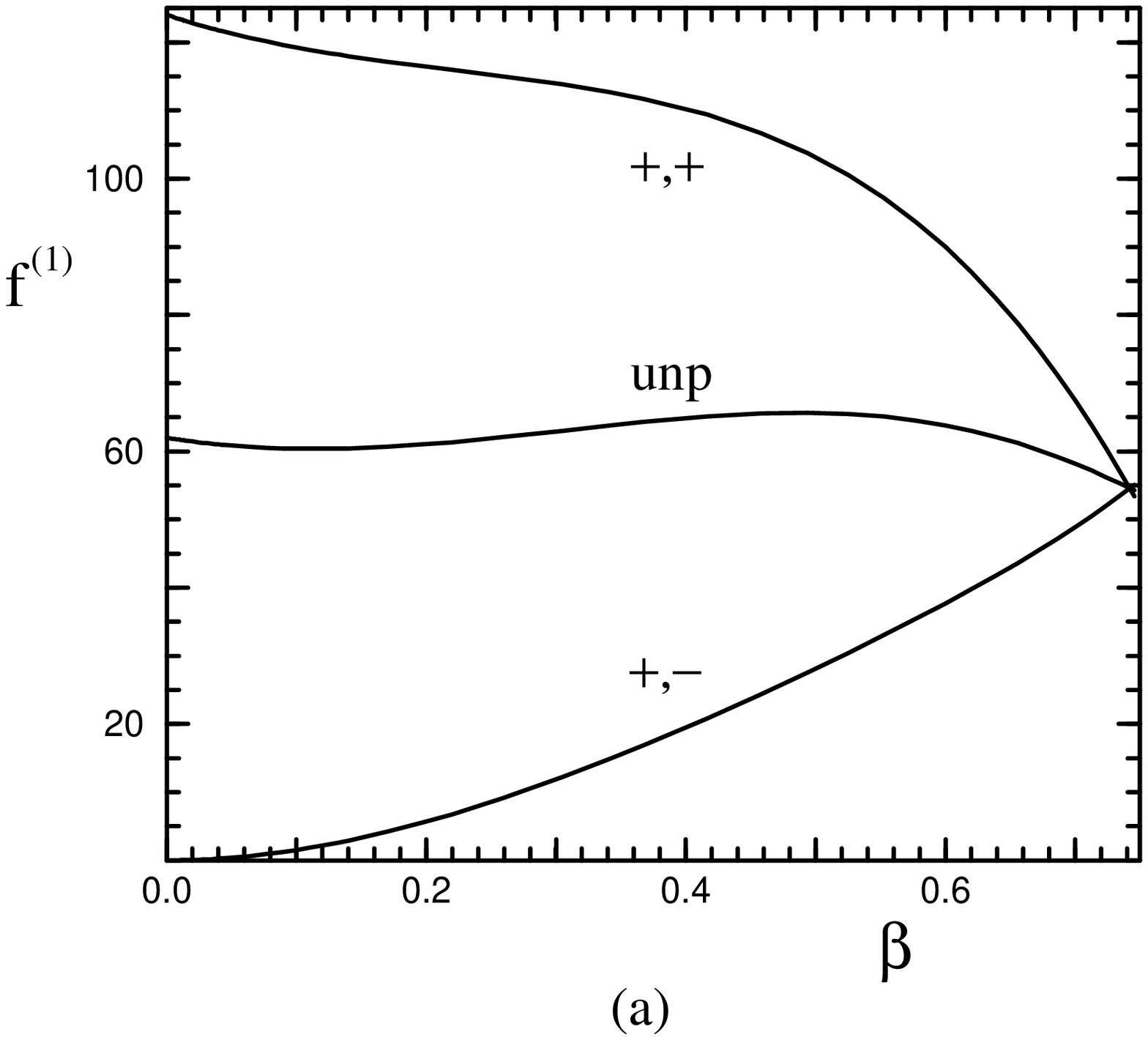,height=6cm,width=8cm}}
%\begin{center}(a)\end{center}
\centerline{\epsfig{file=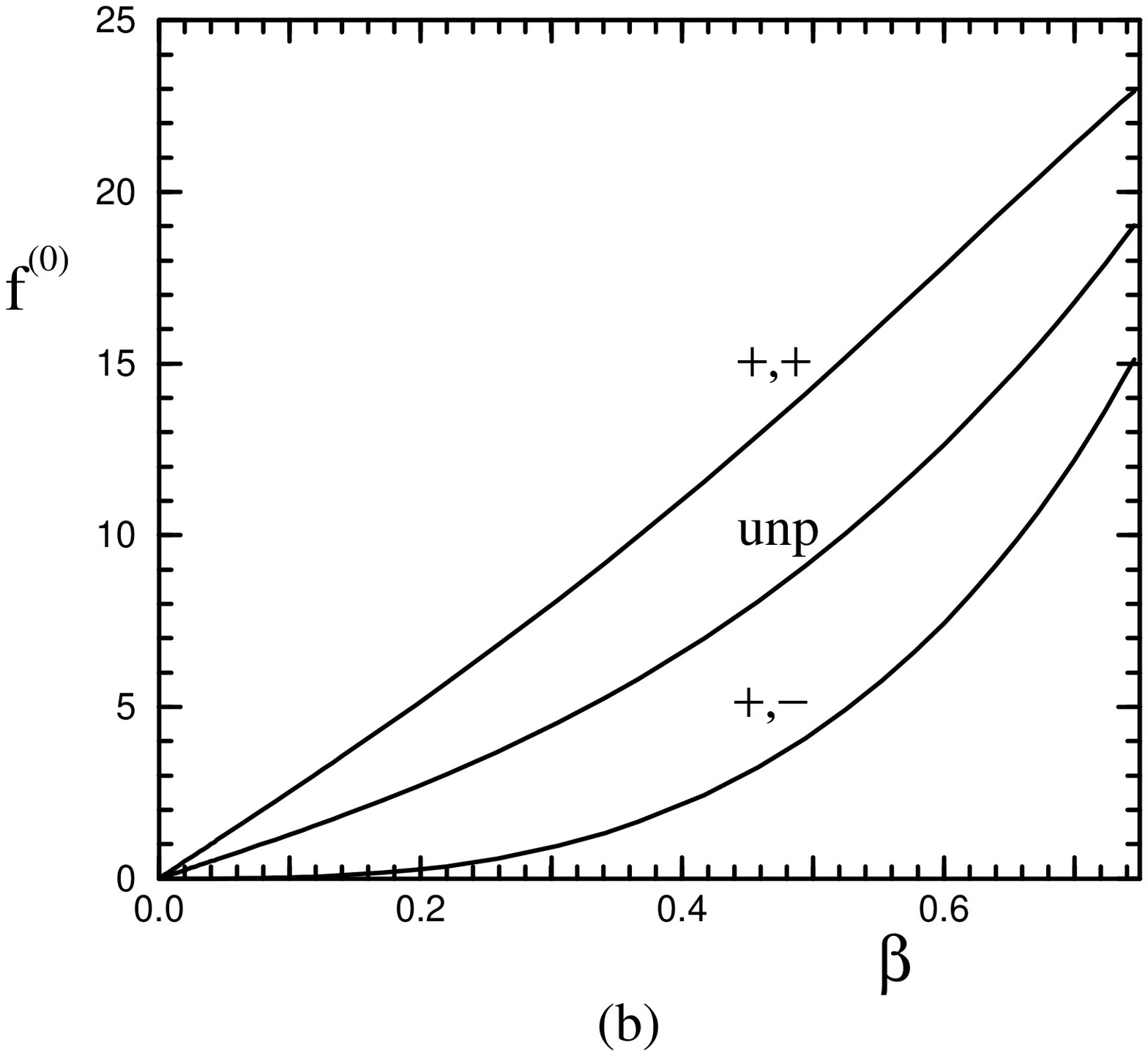,height=6cm,width=8cm}}
%\begin{center}(b)\end{center}
\vspace{10pt}
\caption{The functions (a) $f^{(1)}$; (b) $f^{(0)}$, versus
$\beta$, in the low energy region, for the various helicity states.
}
\label{Fig1}
\end{figure}

\begin{figure}
\centerline{\epsfig{file=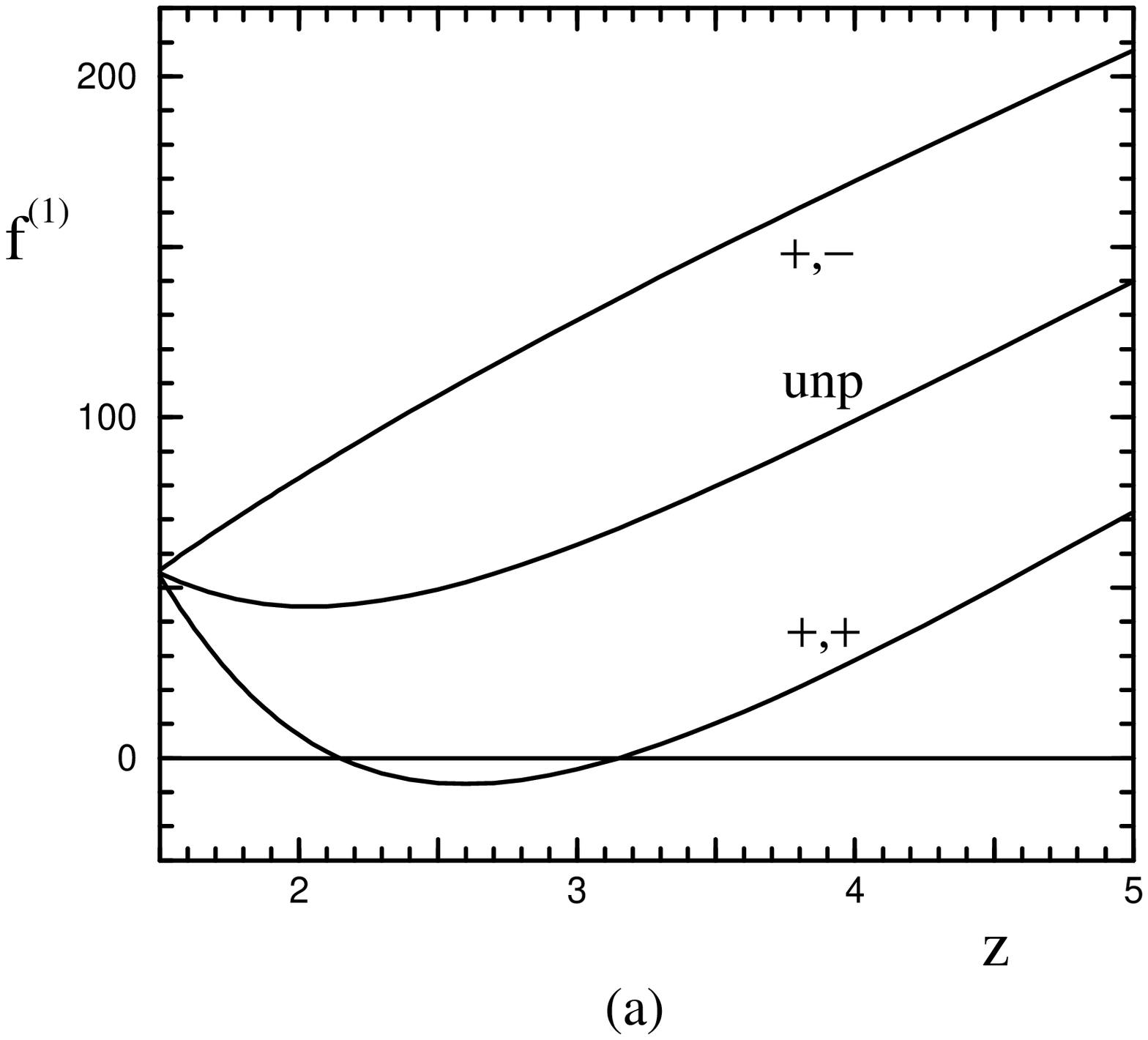,height=6cm,width=8cm}}
%\begin{center}(a)\end{center}
\centerline{\epsfig{file=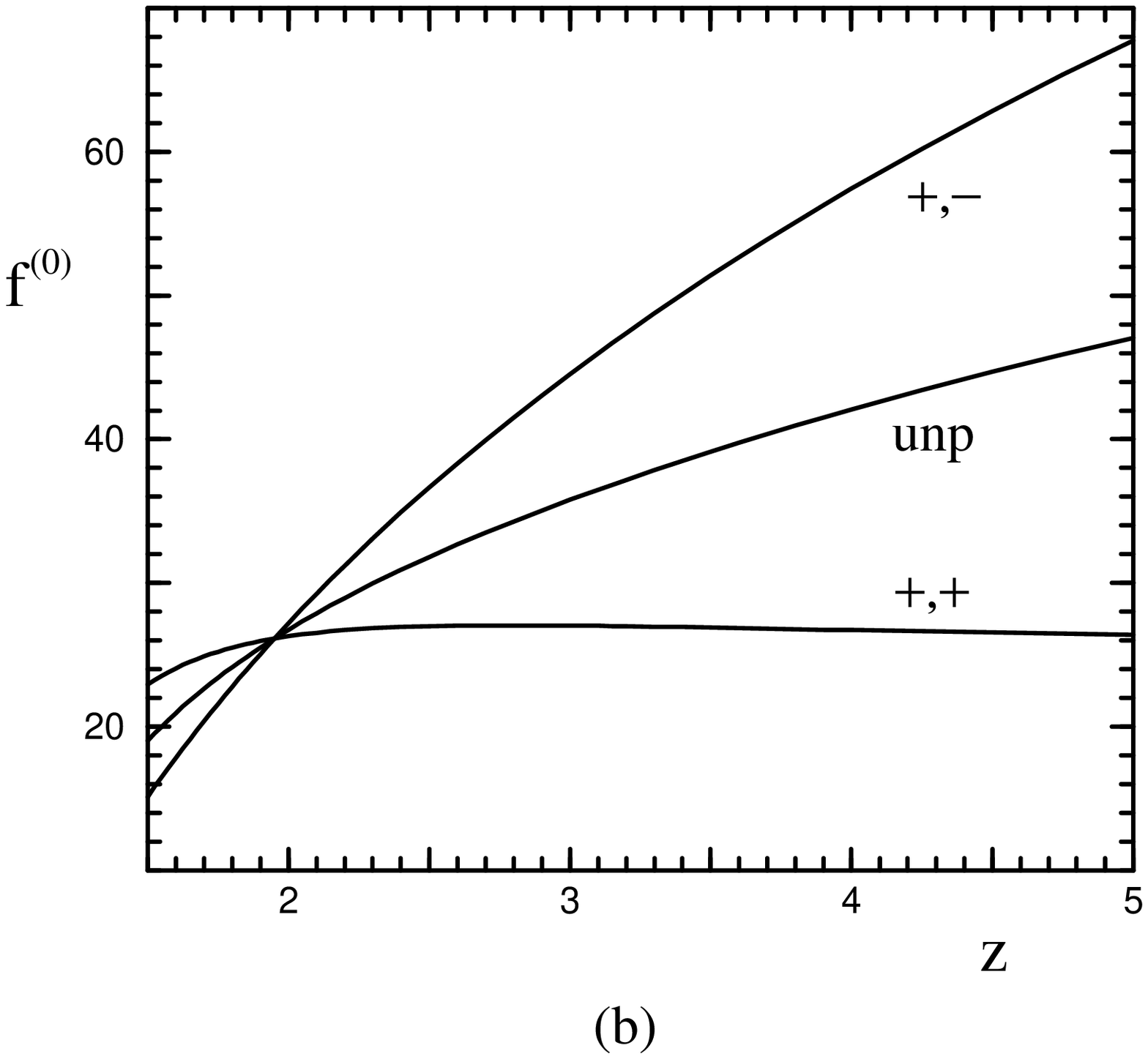,height=6cm,width=8cm}}
%\begin{center}(b)\end{center}
\vspace{10pt}
\caption{The functions (a) $f^{(1)}$; (b) $f^{(0)}$, versus
$z$, in the intermediate energy region, for the various helicity states.
}
\label{Fig2}
\end{figure}

\begin{figure}
\centerline{\epsfig{file=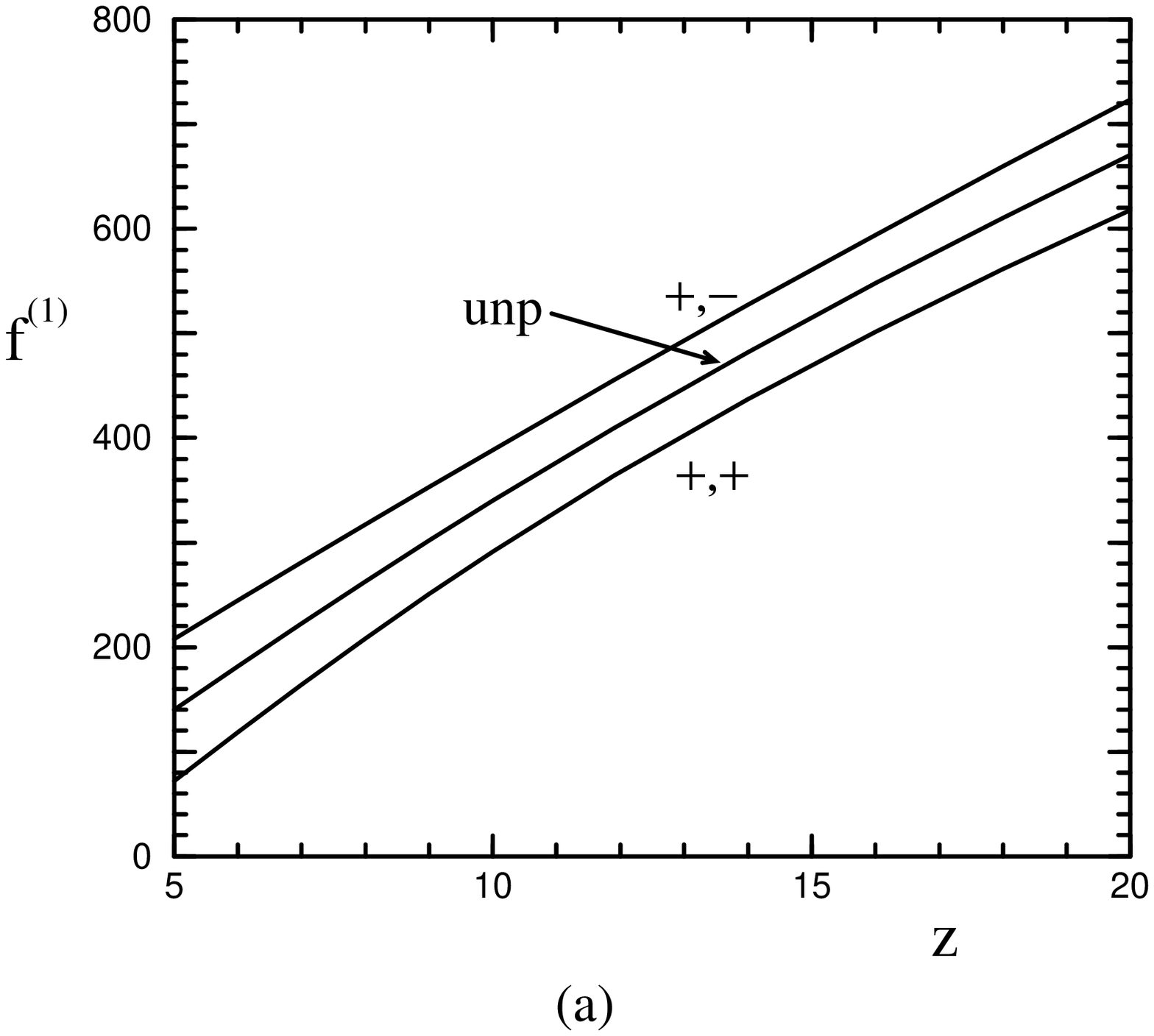,height=6cm,width=8cm}}
%\begin{center}(a)\end{center}
\centerline{\epsfig{file=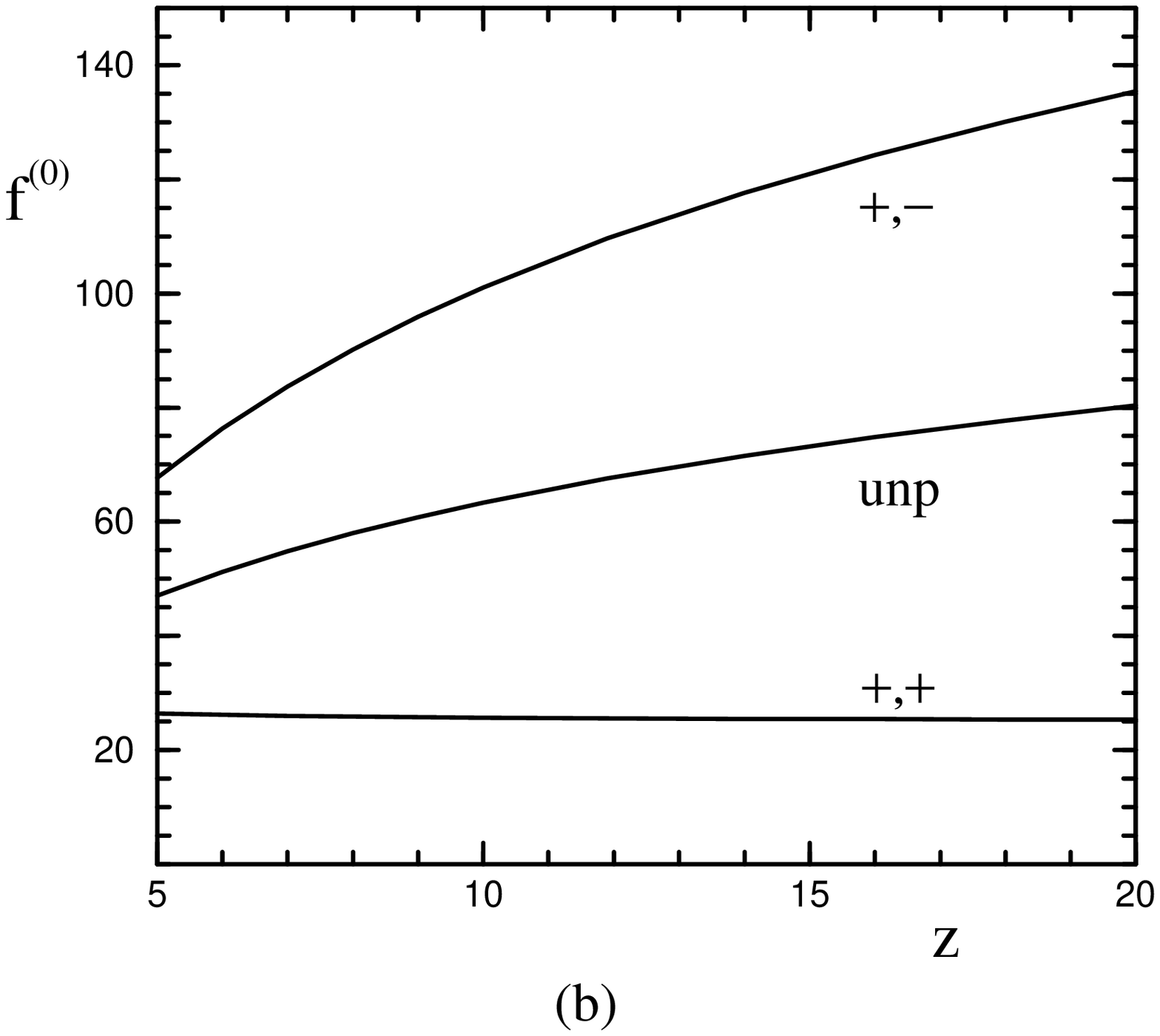,height=6cm,width=8cm}}
%\begin{center}(b)\end{center}
\vspace{10pt}
\caption{The functions (a) $f^{(1)}$; (b) $f^{(0)}$, versus
$z$, in the high energy region, for the various helicity states.
}
\label{Fig3}
\end{figure}

The (exact) plots of $f^{(1)}$, $f^{(0)}$ in the three energy ranges
are given in Figures \ref{Fig1}--\ref{Fig3}. In Fig.\ \ref{Fig1} we
plot $f^{(1)}$, $f^{(0)}$ in the low energy region versus $\beta$.
$\beta$ is more sutiable than $z$ in this region since the 
threshold region becomes compressed and $f^{(1)}$ varies quite rapidly
with $z$ right at threshold. Fig.\ \ref{Fig1} (a) highlights the fact
that $f^{(1)}$ is most naturally decomposed into $f^{(1)}(+,+)$ and
$f^{(1)}(+,-)$ since $f^{(1)}(+,+)$ is monotonically decreasing in the
threshold region while $f^{(1)}(+,-)$ is monotonically increasing.
$f^{(1)}_{\rm unp}$ on the other hand exhibits a rather sudden
dip and peak which seems unnatural, until broken down into 
$(+,+)$ and $(+,-)$ components. From Fig.\ \ref{Fig1} (b) we see that
$f^{(0)}$ is monotonically increasing in this region for both helicity
states.

In Fig.\ \ref{Fig2} we plot $f^{(1)}$, $f^{(0)}$ versus $z$ in the
intermediate energy region. We notice that the $f^{(1)}$ cross just
before $z=1.5$ and the $f^{(0)}$ cross just before $z=2$.
$f^{(1)}(+,+)$ becomes negative and reaches a minimum near $z=2.6$ and
then monotonically increases, as does $f^{(1)}(+,-)$ throughout.
$f^{(0)}(+,-)$ continues to grow while $f^{(0)}(+,+)$ levels off
in accord with (\ref{ef47}). In Fig.\ \ref{Fig3} we plot 
$f^{(1)}$, $f^{(0)}$ in the high energy region. The behaviour remains
unchanged.

%\vglue 1cm
%\begin{center}\begin{large}\begin{bf}
\section{PRECISION $\al_s$ DETERMINATION FROM TOP-QUARK PRODUCTION}
%\end{bf}\end{large}\end{center}
%\vglue .3cm

A high energy $\gm\gm$ collider can be used as a ``factory'' for many
interesting particles: Higgs bosons, $W^\pm$ bosons, top quarks etc...
The beam polarization we be useful in producing Higgs bosons and
reducing $Q\ovl{Q}$ backgrounds. More specifically, the $j=0$ channel
will be of interest. This channel also turns out to be
the channel of interest when trying to determine $\alpha_s$ via top quark
production, making it complementary to the Higgs studies. The reason
is that the cross section, and QCD corrections, are enhanced in this
channel, thereby improving the statistics and the determination of
$\alpha_s$, to which the cross section will be quite sensitive. The
process $\gm\gm \RA t\B{t} + X$ is more powerful than
$e^+e^-\RA Q \ovl{Q} + X$ in determining $\alpha_s$ because the QCD
corrections are quite small in the latter, thus requiring an 
unreasonably large number of events for high precision; the corrections are
suppressed by $\alpha_s/\pi \simeq 4$\%, relative to the Born term.
In $\gm\gm \RA t\B{t} + X$, we can ``pick'' our QCD correction by
choosing the appropriate beam energy. Of course, as one gets too close
to threshold, the perturbation series cannot be trusted, for reasons we
will discuss below. Hence there are limitations.

\begin{figure}
\centerline{\epsfig{file=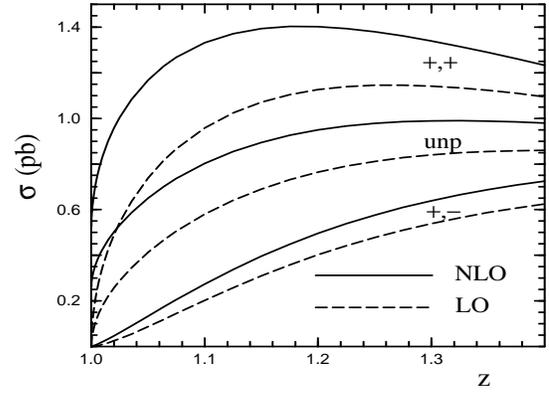,height=6cm,width=8cm}}
\vspace{10pt}
\caption{The $\gm\gm\RA t\B{t}+X$ cross section at LO and NLO, 
versus $z$, for the various helicity states.
}
\label{Fig4}
\end{figure}

To best illustrate the above idea, in Fig.\ \ref{Fig4} we have plotted
the $\gm\gm \RA t\B{t} + X$ cross section at LO and NLO, in the
region $1\leq z \leq 1.4$, for the various helicity states. We took
$N_f= 5$ , $m_t= 174$ GeV  and used $\Lambda = 230$ MeV in the
two-loop expression for $\alpha_s$, evaluated at $\mu^2=s$.  
One could also use $N_f=6$, but since we are not far above threshold
it is simpler to use $N_f=5$ for evolution from $\mu^2=M_Z^2$ to
$\mu^2=s$. 
We notice
that the $j=0$ cross section is the largest, as are its QCD corrections,
in this region. The region $z\simeq 1.2$ is nice in that the $j=0$
cross section is near its maximum and the QCD corrections are 
sizable ($\simeq$ 20\% of the total cross section), 
yet not so large that the perturbative
expansion is unreliable. As one gets closer to threshold, other higher 
order effects, nonperturbative effects and top width effects
may also become important. For these,
and other reasons to be considered below, we will suggest $z=1.2$ as
being the optimal region for extracting $\alpha_s$ and we will give
a rough estimate of how precisely $\alpha_s$ may be determined there.
As well, we suggest the $j=0$ channel as being the most powerful.

Firstly, we note that $z=1.2$ corresponds to $\sqrt{s_{\gm\gm}}\simeq
420$ GeV, for top quark production. This energy should be accessible at
a $\sqrt{s_{e^+e^-}}\gtrsim 500$ GeV NLC. A typical $\gm\gm$ luminosity
assumed is 20 fb$^{-1}$. Since $\sg\simeq 1.4$ pb, this corresponds to
roughly 28,000 $t\B{t}$ events. Since the QCD correction is $\sim 20$\%
of the total cross section,
this translates to $\Dl \alpha_s/\alpha_s\simeq 3\%$, statistically.
With a luminosity increase and, possibly, extended running, one could
envision going to the percent level or better.

The above analysis was purely based on statistics and one-loop QCD 
corrections. Therefore, we will briefly discuss various theoretical
systematic uncertainties. Clearly, one needs a two-loop analysis when
dealing with one-loop corrections of order 20\%, in order to determine
$\alpha_s$ at the level of a few percent. Threshold resummation can
also be performed. One should also take into account the one-loop
electroweak corrections \cite{Denn}. The QED ones are identical in form
to the QCD ones, with the appropriate change in normalization, given
by (\ref{eg6}). There will be a minor dependence on $m_t$, which will
be lessened with future Fermilab runs. The uncertainty on $m_t$
translates to an uncertainty on $z$. Since the $j=0$ cross section
is near its peak for $z\simeq 1.2$, minor variations in $z$ will not
appreciably affect the results. 

Of some concern are resolved photon contributions, where a gluon or 
quark within the photon can participate directly in the interaction.
Suppression of these contributions is a major reason for working
close to threshold. Since the parton distributions within the photon
drop steeply with increasing momentum fraction, $x$, and since $x$
must be large near threshold, such contributions are quite suppressed.
Confirmation of this assertion may be inferred from the resolved 
contributions to $b$ quark production near threshold presented in 
\cite{Drees} from which we conclude that only very poor knowledge (if any)
of the photon structure will be required, as such contributions will
be a fraction of a percent of the cross section. One can further
reduce those contributions by identifying outgoing jets collinear with
one of the photon beams, which are a signature of resolved photon events.
One can also require that the energy deposited in the detectors be 
equal to the total beam energy in order to account for missed jets of the
type mentioned above.

From the experimental side, we are assuming only that $t\B{t}$ events
can be clearly identified. With experience gained from Fermilab, this
seems reasonable, especially considering the cleaner initial and final
states in the $\gm\gm$ case. Another experimental issue is that of
normalization. In order to avoid normalization uncertainties, arising
from luminosity uncertainties, we suggest the measurement of a ratio
of cross sections denoted
\beq
R^{\gm\gm}_{Q/P} \equiv \fr{\sg(\gm\gm\RA Q\ovl{Q}+X)}
{\sg(\gm\gm\RA P\ovl{P}+X)},\SSP  P=W,l.
\eeq
The ratio of $t\B{t}$ to $W^+W^-$ events is statistically quite
powerful as over one million $W^+W^-$ events are expected 
at such a ``W factory''
\cite{JikW}. This highlights
the complementary nature of top quark and $W^\pm$ production at a
$\gm\gm$ collider. As well, electroweak corrections to $W^+W^-$ production have
been studied \cite{JikW}. 
For the same reasons as for $t\B{t}$ production, the
resolved photon contributions will be suppressed. 
If a $b\B{b}$  pair is produced in conjunction with the $W^+W^-$, this
will constitute a background to $t\B{t}$ production.

It is worth discussing the many advantages of determining $\alpha_s$
via $\gm\gm\RA t\B{t}+X$ relative to  some of
the options currently being used.
The calculation is perturbative and avoids nonperturbative
contributions arising in $\alpha_s$ determinations from mass splittings
and tau decays. Other 
determinations, based on evolution of hadronic
structure functions, rely on the parton model and assumed knowledge
of hadronic structure. No such assumptions are made here. Unlike the
3- to 2-jet ratio from $e^+e^-$ annihilation, we avoid having to define
the jet isolation criteria by measuring the total $t\B{t}$ cross section.
Since we are at a large energy scale, not only does perturbation theory
work well, but we automatically determine $\alpha_s$ at (or above) the
$t\B{t}$ threshold, without having to perform evolution
or cross flavor thresholds. From a theoretical
viewpoint, the most comparably clean determination comes from the ratio
of hadrons to lepton pairs produced in $e^+e^-$ annihilation at the
$Z$ pole. As mentioned earlier, the small QCD correction proves an
insurmountable limiting factor in that case.

At this stage, our enthusiasm is dampened somewhat however by the need
for a two-loop calculation. This need is highlighted by the fact that
there is an arbitrariness in the choice of renormalization
scale, $\mu$, which can
only be compensated by the inclusion of two-loop corrections. The variation
of $\alpha_s$ with $\ln\mu$ is order $\alpha_s^2$ though, so for a 
reasonable choice of $\mu$ (i.e.\ $\sqrt{s}$, $m_t$, \ldots) the 
two-loop scale dependent contribution should not be too large and should
not change the value of $\alpha_s$ radically. Nonetheless, as pointed out
earlier, a two-loop calculation will eventually be required. In light
of that fact, we see the importance of having simple analytical 
results for the one-loop corrections as they will be incorporated
in the two-loop result. Also, we do not suggest that one consider this
determination of $\alpha_s$ in isolation. Rather, it should be 
combined with all other precision determinations, including the low
energy ones, in order to minimize the error and provide an excellent
test of QCD at the same time. 

%\onecolumn \noindent

%\vglue 1cm
%\begin{center}\begin{large}\begin{bf}
\section{CONCLUSIONS}
%\end{bf}\end{large}\end{center}
%\vglue .3cm

The analytical results for the one-loop QCD/QED corrections to 
massive fermion production presented in \cite{heavy} were extended by
analytically integrating the single integral (virtual+soft) part.
The differential and integrated cross sections were series expanded
to order $\beta^{10}$ (including $\beta^{11}\ln \beta$ terms) and
were shown to be of practical use as well as being informative.
Accurate parametrizations of the total cross section, valid up to
$\sqrt{s}/2m=20$, were presented. As an application, we showed how
top quark production at a $\gm\gm$ collider capable of reaching
$\sqrt{s}\simeq 420$ GeV could be used to precisely determine
$\alpha_s$, statistics permitting. Theoretical uncertainties were
briefly discussed as were advantages over other $\alpha_s$ determinations.
Those advantages make this method of determination quite appealing.

The importance of performing the two-loop corrections was emphasized.
As well, the best value of $\alpha_s$ will still come from combining
all determinations, including the low energy ones. The major strengths
of this determination are the largeness of the QCD corrections and 
the potential to reduce the theoretical systematic errors via inclusion
of higher order contributions, calculable using well-established perturbation
theory. The major weakness being the tedious nature of the required 
two-loop calculation.

%\vglue 1cm
%\begin{center}\begin{large}\begin{bf}
\section*{ACKNOWLEDGEMENTS}
%\end{bf}\end{large}\end{center}
%\vglue .3cm

We would like to thank S.\ Dawson, W.J.\ Marciano and F.E.\ Paige
for useful discussions and G.\ Ricciardi for help with Mathematica.
One of us (BK) thanks Z.\ Parsa and the 
ITP of UCSB, where part of this work was done, for their hospitality
during the workshop on future high energy colliders. The work of
BK was supported by U.S.\ Department of Energy contract number
DE-AC02-76CH00016.

\onecolumn \noindent

%\widetext 

%\newpage
%\vglue 1cm
%\begin{center}\begin{large}\begin{bf}
\appendix
\section{}
%\end{bf}\end{large}\end{center}
%\vglue .3cm

Here we present the coefficients $a_i(j)$ appearing in the expression
(\ref{ea43}) for the analytically integrated single integral part,
$f^{(1)}_{si}$. They are
\beqa
\nn
a_1 &=&(4 + 14 \beta - 6 \beta^2) (1 + \beta) + j (7 - 9 \beta 
- 19 \beta^2 + 3 \beta^3),
\SSPP
a_2 = 8 [2 \beta^2 (1 + \beta^2)-j (5 + 3 \beta^2)],
\\ \nn
a_3 &=& - 16 + 8 \beta^4 + j (6 - 14 \beta^2),
\SSPP
a_4 = - 8 (1 - \beta^4 + 2 j) (1 + \beta^2),
\\ \nn
a_5 &=& - \{[4 (1 + \beta^2)^2 - \beta (1 - \beta^2)] (1 - \beta^2) +
      j [8 + \beta/2 + 14 \beta^2 - 7 \beta^3 - 2 \beta^4 
+ 5/2 \beta^5]\},
\\ \nn
a_6 &=& (1 - \beta^2)^2 - j/2 (1-14 \beta^2+5 \beta^4),
\SSPP
a_7 = 6 (1 - \beta^2)^2 + j (25-3 \beta+10 \beta^2+\beta^3-9 \beta^4),
\\ \nn
a_8 &=& [(1 - \beta^2)^2+j (15/2+3 \beta^2-5/2 \beta^4)]/2,
\SSPP
a_9 = -(1 - \beta^2)^2+j/2 (1-\beta^2) (17-5 \beta^2),
\\ \nn
a_{10} &=& 4 [(2 - \beta + 2 \beta^2) (1 + \beta)+j (3-19 \beta
-3 \beta^2-7 \beta^3-2 \beta^4)/(3+\beta^2)],
\\ \nn
a_{11} &=& [(1 + \beta^2)^2 - \beta/3 (1 - \beta^2)] ( 1 - \beta^2)+
                j/6 (12+29 \beta
+39 \beta^2-10 \beta^3-9 \beta^4-\beta^5),
\\ \nn
a_{12} &=& \{[(1 + \beta^2) (9 + 9 \beta + 9 \beta^2 + 2 \beta^4)
 - 3 (1 - \beta^2) - 2 \beta^3] (1-  
 \beta) \beta \\ \nn
 && +j/2 (180-219 \beta+165 \beta^2-219
 \beta^3+95 \beta^4-73 \beta^5+35 
                   \beta^6-\beta^7+5 \beta^8)/
  (3+\beta^2)\}/(3+\beta^2), \\ \nn
a_{13} &=&  - [7 (1 - \beta^2)^2+j/2 (73+58 \beta^2-35 \beta^4)]/2,
\SSPP
a_{14} = 8 (1 + \beta^2 -  7/2\, j),
\\ \nn
a_{15} &=& - 16 [1 + \beta^2 -  3j],
\SSPP
a_{16} = - 4 (1 + \beta^2) (1 - \beta^4 + 2 j),
\\ 
a_{17} &=& 8 [1 + 2 \beta^2 - j (19 + 7 \beta^2)/(3+\beta^2)],
\\ \nn
a_{18} &=& 4 [(9+2 \beta^2) (1-\beta^4)-j/2 (33-27 \beta^2-29 
\beta^4-9 \beta^6)/(3+\beta^2)]/(3+\beta^2),
\\ \nn
a_{19} &=& [8 (1+\beta^2)^2-3 \beta (1-\beta^2)] (1-\beta^2)+
   j (16-37/2 \beta+28 \beta^2-11 \beta^3-4 \beta^4+15/2 \beta^5).
\eeqa

%\narrowtext

\section{}

Here we present the $c_i$ entering in the parametrizations for 
$f^{(1)}(+,+)$, $f^{(1)}(+,-)$ in the various energy regions whose
form is given in Equations (\ref{ef67}), (\ref{ef68}), respectively.
For $f^{(1),le}(+,+)$, the coefficients are
\beqa
c_3 &=& -155,\,\,c_4=-125.68,\,\, c_5=119.09, \\ \nn
  c_6 &=& -540.34,\,\,
c_7=364.26\,\,.
\eeqa
For $f^{(1),ie}(+,+)$,
\beqa
\nn
c_0 &=& 53.502,\,\,c_1=-137.56,\,\,c_2=102.57,\,\,c_3=-29.359, \\ 
 c_4 &=& 3.3413,\,\,c_5=.21711,\,\,c_6=-.061446\,\,.
\eeqa
%\twocolumn \noindent
For $f^{(1),he}(+,+)$,
\beqa
c_0 &=& 71.912,\,\,c_1=47.622,\,\,c_2=-.67576,\\ \nn 
 c_3 &=& -2.1675\times 10^{-2},\,\,
c_4=1.1221\times 10^{-3}\,\,.
\eeqa
For $f^{(1),le}(+,-)$,
\beqa
c_5 &=& -667.218,\,\,c_6=2252,\,\,c_7=-5395.85,\\ \nn 
c_8 &=& 8137.55,\,\,
c_9=-6654.48,\,\,c_{10}=2304.95\,\,.
\eeqa
For $f^{(1),ie}(+,-)$,
\beqa
c_0 &=& 55.267,\,\,c_1=57.115,\,\,c_2=-7.3405,\\ \nn 
c_3 &=& 1.3777,\,\,c_4=-.11197\,\,.
\eeqa
For $f^{(1),he}(+,-)$,
\beqa
c_0 &=& 207.66,\,\,c_1=37.016,\,\,c_2=-.15617,\\ \nn 
c_3 &=& -1.2899\times 10^{-3}\,\,.
\eeqa

%\narrowtext
\twocolumn

\newpage

\newpage
\newpage

\newpage
\onecolumn

%\widetext

\begin{table}
\caption{The various $f^{(1)}$ for values of $1.2 \leq z \leq 20$, and
the corresponding single and double integral contributions. Here n.s.\
is the number of significant figures {\em after} the decimal point in
$f^{(1)}$ and $f^{(1)}_{di}$.}

\begin{center}
\begin{tabular}{cccccccc} 
% & \mbox{}  & \mbox{} & \mbox{} & \mbox{} & \mbox{} & \mbox{} & \\ 
 & \mbox{} & $f^{(1)}_{si,{\rm unp}}$ & 
$f^{(1)}_{di,{\rm unp}}$ & 
$f^{(1)}_{\rm unp}$ & $f^{(1)}_{si,{\rm pol}}$ & $f^{(1)}_{di,{\rm pol}}$ 
& $f^{(1)}_{\rm pol}$   \\
  $z$ &   n.s. & $f^{(1)}_{si}(+,+)$ 
 & $f^{(1)}_{di}(+,+)$ & $f^{(1)}(+,+)$
 & $f^{(1)}_{si}(+,-)$ 
 & $f^{(1)}_{di}(+,-)$ & $f^{(1)}(+,-)$
  \\ \hline
\mbox{} & \mbox{} & \mbox{} & \mbox{} & \mbox{} & \mbox{} 
& \mbox{} & \mbox{} \\
 1.2 &  4 & 70.578894 & -5.47998017 & 65.0989 & 33.4162848
   & -1.37661828 & 32.0397
  \\ 
  \mbox{} & 4 & 103.9951788 & -6.85659845 & 97.1386
   & 37.1626092 & -4.10336189 & 33.0592
  \\ 
\mbox{} & \mbox{} & \mbox{} & \mbox{} & \mbox{} & \mbox{} 
& \mbox{} & \mbox{} \\
 2 &  3 & 68.5516 & -24.064 & 44.488 & -76.4447 & 38.792 & -37.653
  \\
 \mbox{} &3 & -7.8931 & 14.728 & 6.835 & 144.9963 & -62.856 & 82.140
  \\ 
\mbox{} & \mbox{} & \mbox{} & \mbox{} & \mbox{} & \mbox{} 
& \mbox{} & \mbox{} \\
 3 & 3 & 92.2075 & -29.594 & 62.614 & -191.7554 & 125.8526 & -65.903
  \\
 \mbox{} &3 & -99.5479 & 96.259 & -3.289 & 283.9629 & -155.447
 & 128.516
  \\ 
\mbox{} & \mbox{} & \mbox{} & \mbox{} & \mbox{} & \mbox{} 
& \mbox{} & \mbox{} \\
 4 &  3 & 132.0495 & -33.0381 & 99.011 & -285.7603 & 215.4483 & -70.312
  \\ 
  \mbox{} &3 & -153.7108 & 182.4102 & 28.699 & 417.8098 & -248.4864
   & 169.323
  \\ 
\mbox{} & \mbox{} & \mbox{} & \mbox{} & \mbox{} & \mbox{} 
& \mbox{} & \mbox{} \\
 5 &  3 & 176.7014 & -36.802 & 139.899 & -367.5540 & 299.902 & -67.652
  \\
 \mbox{} &3 & -190.8526 & 263.100 & 72.247 & 544.2554 & -336.704
   & 207.551
  \\ 
\mbox{} & \mbox{} & \mbox{} & \mbox{} & \mbox{} & \mbox{} 
& \mbox{} & \mbox{} \\
 10 &  3 & 395.3262 & -55.3625 & 339.964 & -688.1880 & 639.5445
   & -48.6435
  \\
 \mbox{} &3 & -292.8618 & 584.182 & 291.320 & 1083.5142 & -694.907
   & 388.607
  \\ 
\mbox{} & \mbox{} & \mbox{} & \mbox{} & \mbox{} & \mbox{} 
& \mbox{} & \mbox{} \\
 20 &  1 & 749.8886 & -79.437 & 670.45 & -1140.3966 & 1086.967 & -53.43
  \\
 \mbox{} &1 & -390.5080 & 1007.530 & 617.02 & 1890.2852 & -1166.404
   & 723.88 
\\ 
\end{tabular} 
\end{center}
\label{TabI}
\end{table}

\begin{table}
\caption{The fractional errors on the various $f^{(1)}$ computed using
the series expansions up to order $\beta^{10}$, for values
of $1.05 \leq z \leq 1.4$.}

\begin{center}
\begin{tabular}{cccccc} 
  $z$ & $\beta$ & ${\rm f.\, err}(+,+)$ &  ${\rm f.\,  err}(+,-)$ & 
${\rm f.\, err}_{\rm unp}$ 
& $\beta^{11}$  
   \\ \hline
\mbox{} & \mbox{} & \mbox{} & \mbox{} & \mbox{} & \mbox{} \\
 1.05 &  .3049 & $2.1\times 10^{-7}$ & $2.5\times 10^{-6}$
 & $4.3\times 10^{-7}$ & $2.1\times 10^{-6}$
  \\ 
 1.1 &  .4166 & $-6.8\times 10^{-6}$ & $2.3\times 10^{-4}$
 & $3.1\times 10^{-5}$ & $6.6\times 10^{-5}$
  \\ 
 1.2 &  .5528 & $-2.0\times 10^{-4}$ & $3.3\times 10^{-3}$
 & $6.9\times 10^{-4}$ & $1.5\times 10^{-3}$
  \\ 
 1.3 &  .6390 & $-1.4\times 10^{-3}$ & $1.3\times 10^{-2}$
 & $3.4\times 10^{-3}$ & $7.3\times 10^{-3}$
  \\ 
 1.4 &  .6999 & $-5.6\times 10^{-3}$ & $2.9\times 10^{-2}$
 & $8.9\times 10^{-3}$ & $2.0\times 10^{-2}$
  \\ 
\end{tabular} 
\end{center}
\label{TabII}
\end{table}


\noindent

\begin{references}
\bibitem{Gun} J.F.\ Gunion and H.E.\ Haber, Phys.\ Rev.\ D {\bf 48},
5109 (1993).
%
\bibitem{heavy}
B.\ Kamal, Z.\ Merebashvili,
and A.P.\ Contogouris,
Phys.\ Rev.\ D {\bf 51}, 4808 (1995); D {\bf 55}, 3229(E) (1997).
%
\bibitem{Jik} G.\ Jikia and A.\ Tkabladze, Phys.\ Rev.\ D {\bf 54},
2030 (1996).
\bibitem{Bord} D.L.\ Borden, V.A.\ Khoze, J.\ Ohnemus, and 
W.J.\ Stirling, Phys.\ Rev.\ D {\bf 50}, 4499 (1994).
\bibitem{Fadin} V.S.\ Fadin, V.A.\ Khoze, and A.D.\ Martin, Phys.\
Rev.\ D {\bf 56}, 484 (1997).
\bibitem{Denn} A.\ Denner, S.\ Dittmaier, and M.\ Strobel, Phys.\
Rev.\ D {\bf 53}, 44 (1996).
\bibitem{Nas} P.\ Nason, S.\ Dawson, and R.K.\ Ellis, Nucl.\
Phys.\ {\bf B303}, 607 (1988); {\bf B327}, 49 (1989);
{\bf B335}, 260(E) (1990).
\bibitem{Been} W.\ Beenakker, H.\ Kuijf, W.L.\ van Neerven, and
J.\ Smith, Phys.\ Rev.\ D {\bf 40}, 54 (1989).
\bibitem{SLAC} D.L.\ Burke et al., Phys.\ Rev.\ Lett.\ {\bf 79},
1626 (1997).
\bibitem{Brink} R.\ Brinkmann et al., DESY-97-048, July 1997, 
hep-ex/9707017.
\bibitem{Kuhn} J.H.\ K\"{u}hn, E.\ Mirkes, and J.\ Steegborn,
Z.\ Phys.\ C {\bf 57}, 615 (1993).
\bibitem{Math} S.\ Wolfram, {\em Mathematica User's Manual},
Addison Wesley, 1991.
\bibitem{REDUCE} A.C.\ Hearn, {\em REDUCE User's Manual Version 3.6}
(Rand Corporation, Santa Monica, CA, 1995).
\bibitem{Lewin} L.\ Lewin, {\em Polylogarithms and Associated Functions},
North Holland, 1981.
\bibitem{Drees} M.\ Drees, M.\ Kr\"{a}mer, J.\ Zunft, and P.M.\
Zerwas, Phys.\ Lett.\ {\bf B306}, 371 (1993).
\bibitem{JikW} G.\ Jikia, Nucl.\ Phys.\ {\bf B494}, 19 (1997).
\end{references}
\end{document}